**MT-LAB**
A VKR CENTRE OF EXCELLENCE

# Roadmap Document on Stochastic Analysis

*Authors:*
Bo Friis NIELSEN
Flemming NIELSON
Henrik PILEGAARD
Michael James Andrew SMITH
Ender YÜKSEL
Kebin ZENG
Lijun ZHANG

October 9, 2018

# Contents

# Preface

This document was prepared as part of the MT-LAB research centre. The research centre studies the Modelling of Information Technology and is a VKR Centre of Excellence funded for five years by the VILLUM Foundation. You can read more about MT-LAB at its webpage www.MT-LAB.dk.

The goal of the document is to serve as an introduction to new PhD students addressing the research goals of MT-LAB. As such it aims to provide an overview of a number of selected approaches to the modelling of stochastic systems. It should be readable not only by computers scientists with a background in formal methods but also by PhD students in stochastics that are interested in understanding the computer science approach to stochastic model checking.

We have no intention of being encyclopedic in our treatment of the approaches or the literature. Rather we have made the selection of material based on the competences of the groups involved in or closely affiliated to MT-LAB, so as to ease the task of the PhD students in navigating an otherwise vast amount of literature.

We have decided to publish the document in case other young researchers may find it helpful. The list of authors reflect those that have at times played a significant role in the production of the document.

# Part I

# Introduction

# Chapter 1

# Introduction

Model checking is a structured approach to system analysis. In this roadmap we put special emphasis on the analysis of transition systems with stochastic features. In particular we focus on systems that can be described by finite Markov chains in discrete or continuous time. In some sense the field has a history that goes back approximately 100 years, however, the structured approach offered by model checking is more recent and has had most of its formulation during the last 20 years.

The roadmap is intended to give the newcomer to the area a fast introduction to the main ideas of the field serving as an introduction to the vast literature in the fields of stochastic model checking and performance evaluation.

In Chapter 2 we give the definition of finite state Markov chains in discrete and continuous time. Markov chains can be used to describe (labelled) transition system. We discuss this in Chapter 3. Chapter 4 contains some important mathematical models useful for many model checking problems and relevant for performance evaluation in general. In Chapter 5 different languages used for model formulation. This is followed by chapter 6 describing enquiry languages, i.e. logical expressions about the models that is to be tested. The actual testing is typical performed using concrete software packages. Some important examples are described in Chapter 7. Not all properties of interest can conveniently be expressed as logical requests. In this case one would term the analysis performance evaluation. This line of thought is pursued in Chapter 8. The distinction between model checking and performance evaluation is somewhat subtle but in general model checking is associated with the structured approach of logical requests to models formulated in a computer algebra.

# Part II

# Fundamental Models in Stochastic Systems

# Chapter 2

# Markov Chains: A Mathematician's Perspective

A Markov chain is a model of a dynamical system with randomness where either the state space or time or both are discrete. We will focus on the discrete state-space. The most important property of a Markov chain is that the part of past behaviour with impact on the future of the process can be summarised in the current state of the process.

We will let $J(t)$ denote the state of the process at time $t$. If $t$ is an element of the integers or another countable set then $J(t)$ can be very general. However, in most cases $J(t)$ will also belong to a countable set, or less frequent to some Euclidean space. If $t$ is an element of the reals then $J(t)$ will belong to some countable set.

## 2.1 Discrete Time and Discrete Space

This is the basic case. The discrete state space can be finite or infinite.

$$P(J(t) = j | J(t-1) = i) = p_{ij}$$

It is customary to collect these probabilities in a matrix $P = \{p_{ij}\}$. The parameters $p_{ij}$ are called the one-stop transition probabilities.

$$\boldsymbol{\pi}(t) = \boldsymbol{\pi}(t-1)P = \boldsymbol{\pi}(0)P^t$$

It is customary to introduce the matrix $P(t) = P^t$, the $(i,j)$th element of which has the probabilistic interpretation $P(J(t) = j | J(0) = i)$. These probabilities are called the $t$–step transition probabilities. Suppose that there is a path of non-zero probability for any state in the state space to any other state. In that case the limit

$$\lim_{t\to\infty} \frac{1}{t} \sum_{k=0}^{t} \boldsymbol{\pi}(t) = \boldsymbol{\pi}$$

exists. There are two possibilities, either $\boldsymbol{\pi}$ is zero or all entries are positive.

## 2.2 Continuous Time and Discrete Space

The majority of continuous time Markov chain models used in computer science have a finite state space. A finite continuous time Markov chain can be viewed as a discrete time Markov chain where the time between state changes is exponentially distributed with a rate that depends on the current state. A continuous time Markov chain is parameterised by its generator matrix $Q$. A generator matrix is characterised by having all row sums being zero and by having non–negative off–diagonal elements. The absolute value of the diagonal element in row $i$ gives the rate of the exponential distribution governing the time spent in state $i$. If state $i$ is absorbing then this rate is 0. The probability transition matrix $P(t)$ is given as a matrix-exponential of the generator matrix

$$P(t) = e^{Qt}$$

where $e^{Qt} = \sum_{k=0}^{\infty} \frac{(Qt)^k}{k!}$. The marginal probabilities $\boldsymbol{p}(t)$ are given by

$$\boldsymbol{p}(t) = \boldsymbol{p}(0) P(t).$$

If there is a path of non-zero probability for any state in the state space to any other state then $\boldsymbol{p}(t) \to \boldsymbol{\pi}$, where $\boldsymbol{\pi}$ can be found as the only non-negative solution to

$$\boldsymbol{\pi} Q = \boldsymbol{0}$$

where the elements of $\boldsymbol{\pi}$ sum to one.

In computer science models one frequently encounter the possibility of events happening that leads back to the originating state. This possibility is not modelled with the generator matrix but can be dealt with in various ways.

# Chapter 3

# Markov Chains: A Computer Scientist's Perspective

In general, *dynamical systems* fall into three categories. *Discrete event systems* are characterised by a discrete but not necessarily finite state space and may, e.g., be described by state transition tables. In contrast, *Continuous state systems* are characterised by a continuous and immediately infinite state space and may, e.g., be described by differential equations. Finally, *hybrid systems* have the characteristics of both discrete event and continuous state systems. In this road-map document we shall be concerned only with discrete event systems, whereas continuous state and hybrid systems are covered by other MT-LAB road-map documents.

A simple way of characterising discrete event systems is afforded by the notion of *transition systems*, i.e. pairs, $T = (S, \longrightarrow)$, consisting of a (possibly infinite) set of *states*, $S$, and a *transition relation*, $\longrightarrow \subseteq S \times S$, defining the possible or allowed set of state changes. In the following we shall write $s \longrightarrow t$ to say that $(s,t) \in \longrightarrow$.

**Example 1 (Transition System)** *The transition system*

$$T = (\{s_1, s_2, s_3\}, \{(s_1, s_2), (s_2, s_3), (s_3, s_1)\})$$

*abstractly characterises a system with three possible states and three transitions allowed between these. The system may be drawn as follows:*

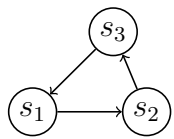

∎

Mostly, however, our interest in systems goes beyond the possible states and transitions in an effort to understand various aspects of the *processes* that may take place in the system. In general, a process may be characterised as a sequence of changes to the *state* of an object. Without loss of generalisation, however, we shall always assume that the object of change is (a part of) the system itself.

In order to fully accommodate our interest in processes we require the means to qualify the actual events that may cause, or be caused by, the system transitions. This requirement leads to the slightly more advanced notion of *labelled transition systems*. A labelled transition system (lts) is a triple, $L = (S, \mathcal{L}ab, \longrightarrow)$, consisting of a set of states, $S$, a set of *labels*, $\ell \in \mathcal{L}ab$ such that $S \cap \mathcal{L}ab = \emptyset$, and a *labelled transition relation*, $\longrightarrow \subseteq S \times \mathcal{L}ab \times S$. Henceforth we shall write $s \xrightarrow{\ell} t$ to mean that $(s, \ell, t) \in \longrightarrow$.

**Example 2 (Labelled Transition System)** *The labelled transition system*

$$L = (\{s_1, s_2, s_3\}, \{a, b\}, \{(s_1, a, s_2), (s_2, b, s_3), (s_3, a, s_2), (s_2, b, s_1)\})$$

*abstractly describes a system with three states, two labels, and four distinct labelled transitions. The system may be drawn as follows:*

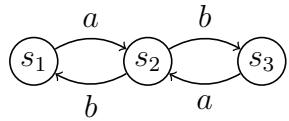

∎

Note that we have carefully avoided giving a clear definition or explanation of the nature of labels. This omission has been made in order ensure the generality of our notion of labelled transition systems. Furthermore, the explained notions of transition systems do not define a start state. This is because the start state is sometimes immaterial to the function of the system and often immaterial to the system or process properties that interest us. Sometimes, however, a particular state will be singled out as the start state of an lts under consideration. We then say that the lts is *rooted*.

## 3.1   Examples

Clearly, the general notion of an lts may be instantiated in many ways. In the following we shall explain how to obtain several well known special cases from various branches of informatics. In the following we shall give some rudimentary such examples.

### 3.1.1 Markovian Processes

**Markov Chains**

First we consider the situation where each event in the system is caused simply by a delay that expires. The underlying notion of time may, of course, be either discrete or continuous; as we shall see, either choice gives rise to a particular model.

Let us first assume that time progresses in discrete steps. At each time step the system changes state in accordance with a transition probability vector, $\vec{p}_i$, that determines the probability, $p_{ij} \in [0,1]$, of a change from the current state, $s_i$, to each state, $s_j$, of the system in a single time-step. Thus we know that $|\vec{p}_i| = |S|$ for all $i$ and we shall further demand that $\sum_{j=1}^{|S|} p_{ij} = 1$ for all $i$ in order for the system to be well-formed. In this situation it is meaningful to choose the set of labels to be the set of permissible probabilities, i.e. $p \in [0,1]$, and the resulting notion of labelled transition system is exactly the class of *Discrete Time Markov Chains* (DTMCs).

**Example 3 (Discrete Time Markov Chain)**

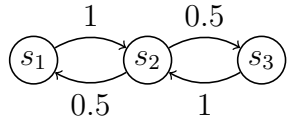

∎

Now let us assume that time progresses continuously. The length of each delay is an exponentially distributed stochastic variable, $X \in \text{Ex}(\lambda)$, where the parameter, $\lambda \in \mathbf{R}$, uniquely determines the underlying exponential distribution. In this situation it is meaningful to choose the set of labels to be the set of permissible parameters, i.e. $\mathcal{L}ab = \mathbf{R}$, and the resulting notion of labelled transition systems is exactly the class of *Continuous Time Markov Chains*.

### 3.1.2 Reactive Systems

Note that, as in Example 4 below, the labels are often related to actions performed by system or user. Therefore the set of labels is sometimes referred to as the set of *actions*, $\alpha \in \mathcal{A}ct$.

**Example 4 (Event Driven State Machine)** *Consider the following labelled transition system:*

$$Lamp = (\{\text{on}, \text{off}\}, \{\text{press}\}, \{(\text{off}, \text{press}, \text{on}), (\text{on}, \text{press}, \text{off})\}),$$

*It abstractly models a lamp. The lamp has two states,* on *and* off*, and it alternates between these state at the* press *of a button, which is the only event recognised by the system. The system may be drawn as follows:*

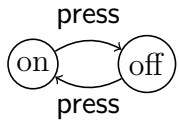

∎

In Computer Science it is customary to think of systems as processes that navigate through a state space in discrete steps prescribed by a set of instructions. Such a system may be defined in any programming language (*syntax*), as long as the instructions of the language have a well-defined effect on the state of the system (*semantics*).

# Part III

# High-Level Modelling

# Chapter 4

# Stochastic Modelling: A Mathematician's Perspective

## 4.1 Notation

In this section, mathematical symbols are involved to denote basic mathematical elements. A matrix is represented by capital primarily roman letters like $\boldsymbol{T}$ and $\boldsymbol{A}$. The symbol $\boldsymbol{I}$ will be used for an identity matrix of appropriate dimension. $\boldsymbol{e}$ is used to denote a column vector of ones of appropriate dimension,

$$\boldsymbol{e} = \begin{bmatrix} 1 \\ 1 \\ \vdots \\ 1 \end{bmatrix}.$$

And $\boldsymbol{0}$ denotes the vector of zeros.

## 4.2 Phase-type Distribution

The phase-type distribution is based on the method of stages, a technique introduced by A.K. Erlang [43] and generalised by Jensen [70] and M.F. Neuts [83]. The key idea is to model random time intervals as being made up of a (possibly random) number of geometric or exponential distributed segments and to exploit the resulting Markovian structure to simplify the analysis. The definition of phase-type distribution by M.F. Neuts [83] is: *a probability distribution on the nonnegative integers is of phase type, if and only if there exists a finite Markov chain with a single absorbing state into which absorption is certain, such that for some choice of the initial probabil-* *ities this distribution is that of the time till absorption. Continuous distribution on [0,∞) of phase type are similarly defined in relation to continuous parameter Markov chains.* In general, many definitions and results regarding discrete time phase-type distribution carry over verbatim to the continuous time case, others need minor modifications.

### 4.2.1 Discrete Phase-type Distribution

A discrete phase-type distribution in a finite discrete time Markov chain with transition matrix $\boldsymbol{P}$ of dimension $m+1$ is given by (4.1). The Markov chain has $m$ transient and 1 absorbing state.

$$\boldsymbol{P} = \begin{bmatrix} \boldsymbol{T} & \boldsymbol{T}^0 \\ \boldsymbol{0} & 1 \end{bmatrix}, \tag{4.1}$$

where $\boldsymbol{T}^0 = (\boldsymbol{I} - \boldsymbol{T})\boldsymbol{e}$.

The initial probability vector is denoted by $(\boldsymbol{\alpha}, \alpha_{m+1})$. The pair $(\boldsymbol{\alpha}, \boldsymbol{T})$ is called a representation of the phase-type distribution. Given a representation of a discrete phase-type distribution, we can calculate the probability mass function by

$$f(x) = \boldsymbol{\alpha}\boldsymbol{T}^{x-1}\boldsymbol{T}^0, \ x > 0. \tag{4.2}$$

Thus, the cumulative distribution function is given by

$$F(x) = 1 - \boldsymbol{\alpha}\boldsymbol{T}^x\boldsymbol{e}, \ x \geq 0. \tag{4.3}$$

The probability generating function of a discrete random variable is a power series representation of the probability mass function of the random variable. The generating function $H(z)$ from a non-negative discrete random variable X is given by

$$H(z) = E(z^X) = \sum_{x=0}^{\infty} z^x f(x), \tag{4.4}$$

where $f(x)$ is probability mass function. For a discrete phase-type random variable, we find

$$H(z) = \sum_{x=0}^{\infty} z^x f(x) = \alpha_{m+1} + \sum_{x=1}^{\infty} z^x \boldsymbol{\alpha}\boldsymbol{T}^{x-1}\boldsymbol{T}^0 = \alpha_{m+1} + z\boldsymbol{\alpha}(\boldsymbol{I} - z\boldsymbol{T})^{-1}\boldsymbol{T}^0. \tag{4.5}$$

From the generating function $H(z)$ we can obtain the factorial moments for a discrete random variable by successive differentiation. For a discrete phase-type variable with representation $(\boldsymbol{\alpha}, \boldsymbol{T})$ we get the factorial moments

$$E(X(X-1)\ldots(X-(k-1))) = k!\boldsymbol{\alpha}(\boldsymbol{I} - \boldsymbol{T})^{-k}\boldsymbol{e}, \quad k \geq 1. \tag{4.6}$$

The matrix $U = (I - T)^{-1}$ is of special importance as the $(i,j)$th element has an important probabilistic interpretation as the expected time spent in state $j$ before absorption conditioned on starting in state $i$.

### 4.2.2 Continuous Phase-type Distribution

A continuous phase-type distribution in a finite continuous time Markov chain with infinitesimal generator matrix $Q$ of dimension $m + 1$ is given by (4.7). The continuous time Markov chain has $m$ transient and 1 absorbing state.

$$Q = \begin{bmatrix} T & T^0 \\ 0 & 0 \end{bmatrix}, \qquad (4.7)$$

where $T^0 = (I - T)e$.

The initial probability vector is denoted by $(\boldsymbol{\alpha}, \alpha_{m+1})$. The pair $(\boldsymbol{\alpha}, T)$ is called a representation of the phase-type distribution. Given a representation of a continuous phase-type distribution, we can calculate the probability density function by

$$f(x) = \boldsymbol{\alpha} e^{Tx} T^0, \; x > 0. \qquad (4.8)$$

Thus, the cumulative distribution function is given by

$$F(x) = 1 - \boldsymbol{\alpha} e^{Tx} e, \; x \geq 0. \qquad (4.9)$$

Let $U = (-T)^{-1}$, then the $(i,j)$th element $u_{ij}$ is the expected time spent in state $j$ given initiation in state $i$ prior to absorption. Thus, we have the first moment of continuous phase-type distribution PH($\boldsymbol{\alpha},T$) as $\boldsymbol{\alpha} U e$, the mean of a distributed random variable.

To calculate either probability density function or cumulative distribution function involves matrix exponential computation (i.e. $e^{Tt}$). There is a very efficient method for the calculation of this matrix-exponential called *uniformization*. Introducing the quantity $\theta = -\min(T_{ii}: 1 \leq i \leq m)$ one rewrites $T = \theta(K-I)$. The matrix $K$ is a sub-stochastic matrix such that $K = I + \theta^{-1} T$. Now

$$e^{Tt} = e^{-\theta t} \sum_{i=0}^{\infty} \frac{(\theta t)^i K^i}{i!}. \qquad (4.10)$$

Formula (4.10) is very well suited for numerical evaluation as all terms in the series are non-negative and since an appropriate level for truncation of the sum can be derived from the Poisson distribution.

### 4.2.3 Closure Properties of Phase-type Distribution

One of the appealing features of either discrete or continuous phase-type distribution is that the class is closed under a number of operations. The closure properties are a main contributing factor to the popularity of phase-type distributions in probabilistic modelling of technical systems. In particular the class is closed under addition, finite mixtures, and finite order statistics.

**Sum of two independent PH variables**

In both discrete and continuous cases, consider two random variables X and Y with representation ($\boldsymbol{\alpha}$,$\boldsymbol{T}$) and ($\boldsymbol{\beta}$,$\boldsymbol{S}$) respectively. Here the dimension of $\boldsymbol{T}$ is $m$ and the dimension of $\boldsymbol{S}$ is $k$. Then the random variable Z = X+Y follows a phase-type distribution with representation ($\boldsymbol{\gamma}$, $\boldsymbol{L}$) given by (4.11).

$$\boldsymbol{L} = \begin{bmatrix} \boldsymbol{T} & \boldsymbol{T}^0\boldsymbol{\beta} \\ 0 & \boldsymbol{S} \end{bmatrix} \qquad (4.11)$$

and $\boldsymbol{\gamma} = (\boldsymbol{\alpha}, \alpha_{m+1}\boldsymbol{\beta})$, $\gamma_{m+k+1} = \alpha_{m+1}\beta_{k+1}$.

**Finite mixtures of phase-type distributions**

In both discrete and continuous cases, given $X_i$ phase-type distributed with representation ($\boldsymbol{\alpha_i}$, $\boldsymbol{T_i}$) we have Z = $I_i X_i$ with $\sum_{i=1}^{k} I_i = 1$ and P($I_i$ = 1) = $p_i$). It is easy to see that the random variable Z is itself phase-type distributed with representation ($\boldsymbol{\gamma}$, $\boldsymbol{L}$) given by (4.12).

$$\boldsymbol{L} = \begin{bmatrix} \boldsymbol{T_1} & 0 & \ldots & 0 \\ 0 & \boldsymbol{T_2} & \ldots & 0 \\ \vdots & \vdots & \vdots & \vdots \\ 0 & 0 & \ldots & \boldsymbol{T_k} \end{bmatrix} \qquad (4.12)$$

and $\boldsymbol{\gamma} = (p_1\boldsymbol{\alpha_1}, p_2\boldsymbol{\alpha_2}, \ldots, p_k\boldsymbol{\alpha_k})$.

**Finite order statistics**

The order statistic of a finite number of independent discrete phase-type distributed variables is itself phase-type distributed. Given X phase-type distributed with ($\boldsymbol{T_x}$, $\boldsymbol{\alpha_x}$) and Y phase-type distributed with ($\boldsymbol{T_y}$, $\boldsymbol{\alpha_y}$) is min(X,Y) phase-type distributed with representation ($\boldsymbol{\gamma}$, $\boldsymbol{L}$) given by

$$\boldsymbol{L} = \boldsymbol{T_x} \otimes \boldsymbol{T_y}, \qquad (4.13)$$

where $\boldsymbol{\gamma} = \boldsymbol{\alpha_x} \otimes \boldsymbol{\alpha_y}$. Further max(X,Y) is phase-type distributed with representation $(\boldsymbol{\gamma}, \boldsymbol{L})$ given by (4.14).

$$\boldsymbol{L} = \begin{bmatrix} \boldsymbol{T_x} \otimes \boldsymbol{T_y} & \boldsymbol{T_x} \otimes \boldsymbol{T_y^0} & \boldsymbol{T_x^0} \otimes \boldsymbol{T_y} \\ 0 & \boldsymbol{T_x} & 0 \\ 0 & 0 & \boldsymbol{T_y} \end{bmatrix} \quad (4.14)$$

and $\boldsymbol{\gamma} = (\boldsymbol{\alpha_x} \otimes \boldsymbol{\alpha_y}, \boldsymbol{\alpha_x}\alpha_{y,m+1}, \alpha_{x,k+1}\boldsymbol{\alpha_y})$. Here the dimension of $\boldsymbol{T_x}$ is $k$ and the dimension of $\boldsymbol{T_y}$ is $m$. We write $\boldsymbol{L^0}$ explicitly:

$$\boldsymbol{L^0} = \begin{bmatrix} \boldsymbol{T_x^0} \otimes \boldsymbol{T_y^0} \\ \boldsymbol{T_x^0} \\ \boldsymbol{T_y^0} \end{bmatrix}$$

In the continuous case, for X ∈ PH($\boldsymbol{T_x}$, $\boldsymbol{\alpha_x}$) and Y ∈ PH($\boldsymbol{T_y}$, $\boldsymbol{\alpha_y}$) min(X,Y) is phase distributed with representation $(\boldsymbol{L}, \boldsymbol{\gamma})$ given by (4.15):

$$\boldsymbol{L} = \boldsymbol{T_x} \otimes \boldsymbol{I_y} + \boldsymbol{I_x} \otimes \boldsymbol{T_y}, \quad (4.15)$$

where $\boldsymbol{\gamma} = \boldsymbol{\alpha_x} \otimes \boldsymbol{\alpha_y}$. And max(X,Y) is phase type distributed with representation $(\boldsymbol{L}, \boldsymbol{\gamma})$ given by :

$$\boldsymbol{L} = \begin{bmatrix} \boldsymbol{T_x} \otimes \boldsymbol{I_y} + \boldsymbol{I_x} \otimes \boldsymbol{T_y} & \boldsymbol{I_x} \otimes \boldsymbol{T_y^0} & \boldsymbol{T_x^0} \otimes \boldsymbol{I_y} \\ 0 & \boldsymbol{T_x} & 0 \\ 0 & 0 & \boldsymbol{T_y} \end{bmatrix} \quad (4.16)$$

and $\boldsymbol{\gamma} = (\boldsymbol{\alpha_x} \otimes \boldsymbol{\alpha_y}, \boldsymbol{\alpha_x}\alpha_{y,m+1}, \alpha_{x,k+1}\boldsymbol{\alpha_y})$, where the dimension of $\boldsymbol{T_x}$ is $k$ and the dimension of $\boldsymbol{T_y}$ is $m$. We give $\boldsymbol{L^0}$ explicitly

$$\boldsymbol{L^0} = \begin{bmatrix} 0 \\ \boldsymbol{T_x^0} \\ \boldsymbol{T_y^0} \end{bmatrix}.$$

### 4.2.4 Non-uniqueness of representations

A main drawback when modelling with phase-type distributions is the non-uniqueness of their representations. In other words, a phase-type distribution is uniquely given by any representation. However, several representations can lead to the same phase-type distribution, similar to the bisimulation relation of state transition systems. The different representations of a phase-type distribution constitute an equivalence relation.

## 4.3 Queueing Theory

Queueing theory [49, 57] involves the mathematical study of queues, or waiting lines. The formation of waiting lines is a common phenomenon that occurs whenever the current demand for a service exceeds the current capacity to provide that service. The ultimate goal of study is to achieve an economic balance between the cost of service and the cost associated with waiting for that service. In real life, queueing systems are surprisingly prevalent in a wide variety of contexts. Four broad classes of queueing systems are commercial service systems, transportation service systems, business-industrial internal service systems, and social service systems. Therefore, queueing theory contributes on many decisions in reality by predicting various characteristics of the waiting line such as the average waiting time.

### 4.3.1 Basic Structure of Queueing Models

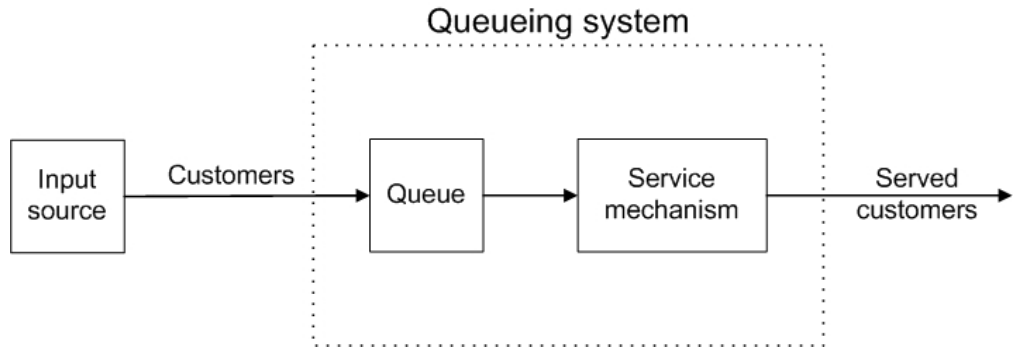

Figure 4.1: Basic structure of queueing models

The basic structure of queueing progress is depicted in Fig. 4.1. *Customers* requiring service are generated over time by an *input source*. These customers enter the *queueing system* and join a (finite or infinite) *queue*. At certain times, a member of the queue is selected for service by some rule known as the *queue discipline* (e.g. first-come-first-served). The required service is then performed for the customer by their *service mechanism*, after which the customer leaves the queueing system.

### 4.3.2 Kendall's notation

In queueing theory, models conventionally are labelled by Kendall's notation for characterising, depicted in fig 4.2.

The most common process is denoted as

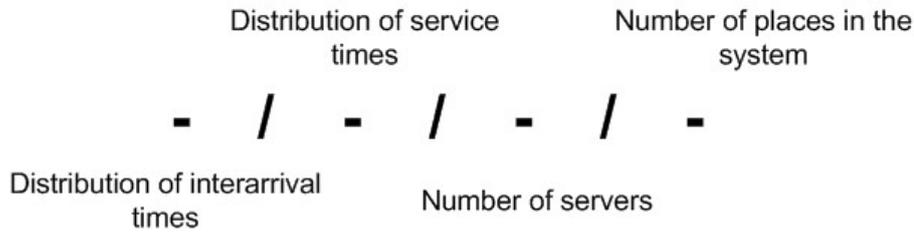

Figure 4.2: Kendall's notation

M = exponential distribution (Markovian)

D = degenerate distribution (constant times)

$E_k$ = Erlang distribution (shape parameter = k)

G = general distribution (any arbitrary distribution allowed)

### 4.3.3 Terminology and Notation

The state of a queueing model is typically defined as the number of customers in the system. In standard terminology, $N(t)$ denotes the number of customers in queueing system at time $t$ $(0 \leq t)$. The system begins in an initial state, and the state of system will change as time elapses. The system is said to be in a *transient condition*. However, after sufficient time has elapsed, the state of the system becomes essentially independent of the initial state and the elapsed time. The system has now essentially reached a *steady-state condition*. Queueing theory has tended to focus largely on the steady-state condition, partially because the transient case is more difficult analytically.

By eliminating the number of customers being served from the state of queueing model, the queue length can be obtained. Some standard quantities considered in queueing theory are

$s$ = number of servers (parallel service channels) in queueing system

$\lambda_n$ = mean service rate of new customers when $n$ customers are in system. In case $\lambda_n$ remains constant, this constant is denoted by $\lambda$.

$\mu_n$ = mean service rate system when there are $n$ customers in the system. *Note* : $\mu_n$ represents combined rate at which all busy servers achieve service completions. When the mean service rate per busy server is constant, this constant is denoted by $\mu$ (In this case, $\mu_n = s\mu$).

$\rho = \lambda/(s\mu)$ is the utilisation factor for the service facility, where it represents the fraction of the system's service capacity ($s\mu$) that is being utilised on the average by arriving customers ($\lambda$). *Note* : $\rho < 1$ is the usual stability criterion in queueing models, where the queue attains an equilibrium as $t \to \infty$. There is no steady-state condition if $\rho \geq 1$.

Steady-state condition

$L$ = expected number of customers in queueing system.

$L_q$ = expected queue length (excludes customers being served).

$W = \text{E}(\omega)$, where $\omega$ = waiting time in system (includes service time) for each individual customer.

$W_q = \text{E}(\omega_q)$, where $\omega_q$ = waiting time in system (excludes service time) for each individual customer.

The relationships between $L$, $L_q$, $W$ and $W_q$ are given as *Little's formula*.

1. $L = \lambda W$
2. $L_q = \lambda W_q$
3. $W = W_q + \frac{1}{\mu}$

*Note* : *Little's formula* is extremely important in a sense that they enable all four of the fundamental quantities to be immediately determined as soon as one is found analytically.

### 4.3.4 Types of Queueing Models

There have been many studies on various types of queueing models. The systems, in order M/M/1, M/G/1, G/M/1 and G/G/1, give the insights from the special queues with Markovian characteristics to greater generality, which is also roughly in order of increasing difficulty. We refer to chapter 11 in [49] and chapter 15 in [57] for further reading.

## 4.4 Renewal Process

A renewal process is a recurrent-event process with independent and identically distributed inter-event times. It is a generalisation of the Poisson process, where the exponential assumption from the Poisson process is relaxed. The asymptotic behaviour of a renewal process is described by the renewal theorem and the elementary renewal theorem.

### 4.4.1 Definition

The general distribution of renewal process is depicted in fig 4.3.

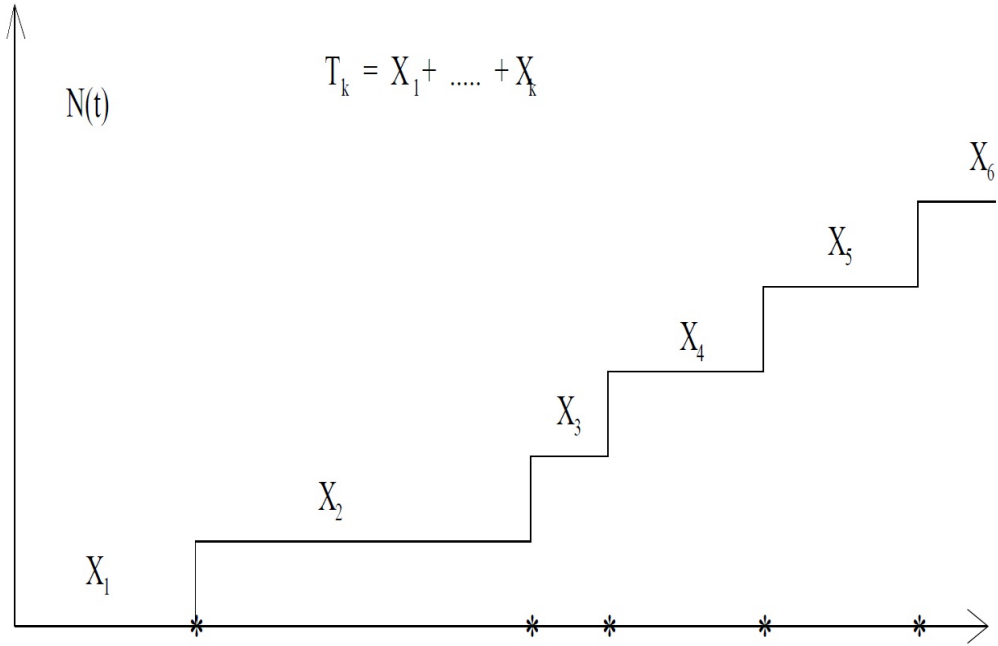

Figure 4.3: A renewal process

$N(t)$: The number of occurrences of some event in the time interval $[0,t]$

$X_i$ : $i$th inter-arrival time

$T_k$ : time of the $k$th arrival

The formal definition of a renewal process $N = \{N(t) : t \geq 0\}$ is a process such that
$$N(t) = max\{n : T_n \leq t\},$$
where $T_0 = 0$, $T_n = X_1 + X_2 + \cdots + X_n$ for $n \geq 1$, and $\{X_i\}$ is a sequence of independent identically distributed positive random variables.

In renewal process, the first interval is allowed to have a different distribution. Let $X_1$ be distributed according to a distribution $H(t)$ and $X_k$ ($k \geq 2$) be distributed according to a distribution $F(t)$. If $H(t) = F(t)$, the process is called a renewal or more standard ordinary renewal process. If $H(t) \neq F(t)$, the process is called a modified or delayed renewal process.

### 4.4.2 Theorems

In renewal theory, study focuses on distribution of $N(t)$ and moments of $N(t)$. A fundamental relationship in renewal process is

$$P\{N(t) < n\} = P\{T_n > t\}.$$

The expected number of events, $E(N(t))$, is given by renewal function, denoted by $m(t)$. Let $F_k$ be the distribution function of $T_k$, renewal function is calculated by

$$m(t) = E(N(t)) = \sum_{n=1}^{\infty} F_n(t).$$

The method of Laplace-Stieltjes transforms (see Definition of Appendix I in [49]) is often useful in renewal theory. The renewal function is expressed as a Laplace-Stieltjes transform

$$m^*(\theta) = \frac{H^*(\theta)}{1 - F^*(\theta)} \quad for\ \theta \neq 0.$$

For $H(t) = F(t)$, standard ordinary renewal process

$$m^*(\theta) = \frac{F^*(\theta)}{1 - F^*(\theta)} \quad for\ \theta \neq 0.$$

For large values of $t$, let $\mu = E(X_i)$ ($i \geq 2$) be the mean of a typical inter-arrival time, then we have the elementary renewal theorem.

$$\frac{1}{t} m(t) \to \frac{1}{\mu} \quad as\ t \to \infty.$$

We refer to chapter 10 in [49] for further reading.

# Chapter 5

# Stochastic Modelling: A Computer Scientist's Perspective

## 5.1 Performance Evaluation Process Algebra (PEPA)

The Performance Evaluation Process Algebra, commonly known as PEPA, was invented by Jane Hillston during her PhD studies, 1991 – 1994, at the University of Edinburgh. The resulting dissertation, published as part of the Distinguished Dissertations in Computer Science series [59], remains the authoritative reference to the calculus. At the time PEPA was not alone but emerged as one of a number of stochastic process algebras (SPAs) developed roughly at the same time [20, 28, 48].

A key motivation for the development of PEPA, and indeed the use of process algebra based paradigms for performance modelling, was *compositionality* [58]. A compositional approach allows complicated systems to be modelled in a systematic manner. Simple low-level components are modelled first and higher-level components are subsequently modelled as compositions of low-level ones. Thus, as in pure process algebra [19, 64, 80], a system is modelled as a complex of interacting *components*. The behaviour of each component is defined by the *actions* that it can perform or as a *composition* of smaller components. In contrast to pure process algebra, though, actions are not assumed instantaneous. Instead, each action is associated with an exponentially distributed random variable that characterises its duration.

The early SPAs all agreed on a core set of algebraic combinators: The *action prefix*, $(\alpha, r).P$, describes a system that is ready to perform action $\alpha$ at the rate $r$ and then go on to behave as described by $P$. The *choice*, $P + Q$, denotes a process that may go on to behave either as described by $P$ or as described by $Q$. The *action hiding* construct, $P/L$, restricts the scope of visibility for the actions mentioned in the set $L$ to cover only $P$. Finally, a notion of *parallel composition* that is based on the CSP concurrency operator [64], $P \bowtie_L Q$ (written $P||_L Q$ in non-PEPA SPAs), denotes a system composed of two processes, $P$ and $Q$, that may go on to unfold their individual behaviour independent of one another except for actions mentioned by the *cooperation set*, $L$, on which the two processes must *synchronise*. Thus the syntax of the early SPAs is generally given by (a variant of) the following grammar for PEPA:

$$P ::= (\alpha, r).P \mid P + Q \mid P \bowtie_L Q \mid P/L \mid A$$

Formally, the behaviour modelled by a PEPA term is a labelled transition system inductively defined by the rules and axioms of the following structural operational semantics:

$$\frac{}{(\alpha, r).P \xrightarrow{(\alpha,r)} P} \qquad \frac{P \xrightarrow{(\alpha,r)} Q}{A \xrightarrow{(\alpha,r)} Q} \text{ if } A \stackrel{\text{def}}{=} P$$

$$\frac{P \xrightarrow{(\alpha,r)} P'}{P + Q \xrightarrow{(\alpha,r)} P'} \qquad \frac{Q \xrightarrow{(\alpha,r)} Q'}{P + Q \xrightarrow{(\alpha,r)} Q'}$$

$$\frac{P \xrightarrow{(\alpha,r)} P'}{P \bowtie_L Q \xrightarrow{(\alpha,r)} P' \bowtie_L Q} \text{ if } \alpha \notin L \qquad \frac{Q \xrightarrow{(\alpha,r)} Q'}{P \bowtie_L Q \xrightarrow{(\alpha,r)} P \bowtie_L Q'} \text{ if } \alpha \notin L$$

$$\frac{P \xrightarrow{(\alpha,r)} P'}{P/L \xrightarrow{(\alpha,r)} P'/L} \text{ if } \alpha \notin L \qquad \frac{P \xrightarrow{(\alpha,r)} P'}{P/L \xrightarrow{(\tau,r)} P'/L} \text{ if } \alpha \in L$$

$$\frac{P \xrightarrow{(\alpha,r_1)} P' \quad Q \xrightarrow{(\alpha,r_2)} Q'}{P \bowtie_L Q \xrightarrow{(\alpha,R)} P' \bowtie_L Q'} \quad \begin{array}{l} \text{if } \alpha \in L \\ \text{and where } R = \frac{r_1}{r_\alpha(P)} \frac{r_2}{r_\alpha(Q)} \min(r_\alpha(P), r_\alpha(Q)) \end{array}$$

Note, that the first eight of these rules are practically shared by all of the early SPAs. The last rule, however, expresses Hillston's view on the nature of synchronisation as encoded in PEPA [58]. Each of the early proposals represented a particular view on this aspect that is formally expressed in the definition of the duration of the delay that occurs when two actions are synchronised. This issue has proven to be a both crucial and controversial aspect of design in stochastic process algebras.

In PEPA the perception of synchronisation is governed by both conceptual and pragmatic concerns. Conceptually PEPA takes the view that action synchronisation represents the cooperation of two equal partners; hence the total time of the event is determined by the slower of the two partners [59]. Pragmatically PEPA insists that rather than just a labelled transition system the behaviour of a model must be a discrete time Markov chain. The latter of these concerns dictates that the resulting duration must be characterised by an exponentially distributed random variable. The former concern requires that the rate, $R$, that characterises the distribution must be related to the minimum of the rates, $r_1$ and $r_2$, that characterises the delays of the individual partners.

These concerns motivated Hillston to define the duration of the resulting delay as an exponentially distributed random variable characterised by the rate, $R = \frac{r_1}{r_\alpha(P)} \frac{r_2}{r_\alpha(Q)} \min(r_\alpha(P), r_\alpha(Q))$. Here $r_\alpha(P)$ denotes the so-called *apparent rate* of the action $\alpha$ in the sub-process $P$, i.e. the sum of the rates of all occurrences of $\alpha$ in $P$ that might compete to participate in the interaction. So $\frac{r_1}{r_\alpha(P)}$ denotes the probability that the action $(\alpha, r_1)$ is the participant from $P$, $\frac{r_2}{r_\alpha(Q)}$ is the similar probability of $(\alpha, r_2)$ being the participant from $Q$, and $\min(r_\alpha(P), r_\alpha(Q))$ selects the lowest apparent rate as a base rate, thereby effectively expressing the idea that the slowest participant decides the duration. Note, that situations where one participant is active and the other passive is easily modelled in this framework by allowing $\infty$ as a rate for some actions.

Over the years Hillston's choices have been subject to some debate. On the one hand, the main point of the critics is that the class of exponential distributions is not closed under maximum; hence PEPA's characterisation of the duration of synchronisation cannot be accurate in all situations and must be considered a pragmatic approximation. On the other hand, the main argument of the proponents is that the computational advantages of Markov chains over more general representations outweighs the imprecision of the approximation.

Regardless of the debate, Hillston's ideas have arguably been very influential and a number of other process algebras have adopted her notion of synchronisation [86, 87]. The PEPA calculus has also been very successful from a practical point of view and a number of tools now support PEPA [24, 32, 46, 63, 94]. In terms of usage PEPA has by now proven its value in a number of contexts, such as traditional performance modelling [40, 47, 62, 65], network security [25, 27, 97, 98], and systems biology [29, 30, 44].

## 5.2 Interactive Markov Chains (IMC)

Interactive Markov Chains (IMC) was developed by Holger Hermann's during the 1990'es. The resulting algebraic formalism is perhaps best described by [26]. As the language is very expressive the standard notion of Markov Chains is not a sufficient semantic model. For this reason Hermann's Phd dissertation, published as a monograph in the series of Springer Lecture Notes [55], takes the position that the Interactive Markov Chain is itself a fundamental (semantic) model, related to Markov Decision Processes and Markov Automata [41, 67, 71], accompanied by an algebraic syntax. In the following we shall assume the view of [26] and present IMC as a process algebraic formalism that requires a rich semantic model.

The fundamental goal of IMC is to provide a *compositional methodology* for modelling and analysis with Markov Chains [55]. Like other SPAs the language relies on traditional process algebraic notions and primitives in order to achieve this goal. Thus, we assume a countably infinite set of *action labels*, $a \in \mathcal{A}$, denoting visible actions as well as a single distinguished internal action, $\tau$, and write $\alpha \in \mathcal{A}_\tau = \mathcal{A} \cup \{\tau\}$ to denote actions in general. Further, we assume a countably infinite set of *process variables*, $X \in \mathcal{V}$. In this context the following syntax describes the (sequential) process algebraic fragment of IMC:

$$\mathcal{P} \ni P ::= \mathbf{0} \mid \alpha.P \mid P + Q \mid X \mid [X := P]_i \mid \ldots$$

As customary for process algebraic languages, $\mathbf{0}$, is used to denote the *terminal process* that is stuck and can perform no further actions.

In contrast, the *action prefix*, $\alpha.P$, describes a process that is ready to perform action $a$ and then go on to behave as described by $P$.

The *choice*, $P + Q$, denotes a process that may go on to behave either as described by $P$ or as described by $Q$.

Finally, the shorthand notation $[X := P]$ is used for an arbitrary (finite) set of defining equations of the form

$$\begin{bmatrix} X_1 := P_1 \\ \vdots \\ X_n := P_n \end{bmatrix}$$

with $X_j \in \mathcal{V}$, and $P_j$ complying to the above grammar. These equations denote a set of *mutually recursive processes*, where each $X_j$ denotes an entry point that is essentially just a named internal state. The subscript $i$ is used to denote the currently active such internal state (or equation).

It is fundamentally assumed that actions take no time, hence the expressions of this fragment simply denote labelled transition systems in accordance with the following structural operational semantics:

$$\frac{}{\alpha.P \xrightarrow{\alpha} P} \qquad \frac{P_i\{[X:=P]_j/X_j\} \xrightarrow{\alpha} P'}{[X:=P]_i \xrightarrow{\alpha} P'}$$

$$\frac{P \xrightarrow{\alpha} P'}{P+Q \xrightarrow{\alpha} P'} \qquad \frac{Q \xrightarrow{\alpha} Q'}{P+Q \xrightarrow{\alpha} Q'}$$

In order to facilitate modelling with Markov chains the syntax is enriched with the *delay prefix*, $(\lambda).P$, that describes a process that is ready to delay for an amount of time that is exponentially distributed with parameter $\lambda \in \mathbb{R}$ and then go on to behave as $P$ once the delay expires.

The additional expressiveness offered by this extension is obvious from the fact that any Markov chain can be modelled by the following syntactic fragment:

$$\mathcal{P} \ni P ::= \mathbf{0} \mid (\lambda).P \mid P+Q \mid X \mid [X:=P]_i \mid \ldots$$

The correspondence between such algebraic expressions and Markov chains is captured by the following structural operational semantics:

$$\frac{}{(\lambda).P \dashrightarrow{\lambda} P} \qquad \frac{P_i\{[X:=P]_j/X_j\} \dashrightarrow{\lambda} P'}{[X:=P]_i \dashrightarrow{\lambda} P'}$$

$$\frac{P \dashrightarrow{\lambda} P'}{P+Q \dashrightarrow{\lambda} P'} \qquad \frac{Q \dashrightarrow{\lambda} Q'}{P+Q \dashrightarrow{\lambda} Q'}$$

As already pointed out, any action labelled transition system, often called a *reactive* or *interactive system*, can be modelled in the process algebraic fragment of this language. As is usually the case for such systems $\xrightarrow{\alpha}$ is simply an ordinary binary relation for any distinct $a$ because $\alpha.P + \alpha.P \approx \alpha.P$. Similarly, any Markov chain can be described using the Markovian fragment. In this case, however, an ordinary relation does not suffice because $(\lambda).P + (\lambda).P \approx (2\lambda).P$. In order to accommodate this we have to accept that $\dashrightarrow{\lambda}$ is a binary *multi-relation* for any distinct $\lambda$.

When the full language is used, however, actions and delays can be interleaved in arbitrary manners. Thus the resulting models are neither reactive/interactive systems or Markov chains. Instead we obtain an *interactive Markov chain* – a model as expressive as the continuous time Markov decision process but notationally more convenient. In particular, the use of action labels allows the sought after compositionality to be realised using the two final language primitives:

$$\mathcal{P} \ni P ::= \text{hide } a_1 \cdots a_k \text{ in } P \mid P\,|[a_1 \cdots a_k]|\,Q \mid \ldots$$

The *parallel composition*, $P\,|[a_1 \cdots a_k]|\,Q$, denotes a system composed of two processes, $P$ and $Q$, that may go on to unfold their individual behaviour independent of one another except for actions mentioned by the *interaction set*, $a_1 \cdots a_k$, on which the two processes must *synchronise*.

The *action hiding* construct, hide $a_1 \cdots a_k$ in $P$, restricts the scope of visibility for the actions mentioned in $a_1 \cdots a_k$ to cover only $P$. This is used to control the synchronisation structure.

Semantically, the composition of two IMCs gives rise to a new IMC in accordance with the following rules of structural operational semantics:

$$\frac{P \xrightarrow{\alpha} P'}{P\,|[a_1 \cdots a_k]|\,Q \xrightarrow{\alpha} P'\,|[a_1 \cdots a_k]|\,Q} \quad \alpha \notin a_1 \cdots a_k$$

$$\frac{Q \xrightarrow{\alpha} Q'}{P\,|[a_1 \cdots a_k]|\,Q \xrightarrow{\alpha} P\,|[a_1 \cdots a_k]|\,Q'} \quad \alpha \notin a_1 \cdots a_k$$

$$\frac{Q \xrightarrow{\alpha} Q' \quad P \xrightarrow{\alpha} P'}{P\,|[a_1 \cdots a_k]|\,Q \xrightarrow{\alpha} P'\,|[a_1 \cdots a_k]|\,Q'} \quad \alpha \in a_1 \cdots a_k$$

$$\frac{P \xrightarrow{\alpha} P'}{\text{hide } a_1 \cdots a_k \text{ in } P \xrightarrow{\alpha} \text{hide } a_1 \cdots a_k \text{ in } P'} \quad a \notin a_1 \cdots a_k$$

$$\frac{P \xrightarrow{\alpha} P'}{\text{hide } a_1 \cdots a_k \text{ in } P \xrightarrow{\tau} \text{hide } a_1 \cdots a_k \text{ in } P'} \quad a \in a_1 \cdots a_k$$

$$\frac{P \dashrightarrow^{\lambda} P' \quad Q \not\xrightarrow{\tau}}{P\,|[a_1 \cdots a_k]|\,Q \dashrightarrow^{\lambda} P'\,|[a_1 \cdots a_k]|\,Q}$$

$$\frac{Q \dashrightarrow^{\lambda} Q' \quad P \not\xrightarrow{\tau}}{P\,|[a_1 \cdots a_k]|\,Q \dashrightarrow^{\lambda} P\,|[a_1 \cdots a_k]|\,Q'}$$

$$\frac{P \dashrightarrow^{\lambda} P' \quad \text{hide } a_1 \cdots a_k \text{ in } P \not\xrightarrow{\tau}}{\text{hide } a_1 \cdots a_k \text{ in } P \dashrightarrow^{\lambda} \text{hide } a_1 \cdots a_k \text{ in } P'}$$

Like PEPA the language of Interactive Markov Chains is very much influenced by the multi-way synchronisation paradigm of CSP. Hence, the first five of the above rules encode this synchronisation paradigm and are more or less standard in the CSP school of process algebra. In adopting these rules IMC further encodes the unique view on delays that separates it from other SPAs, e.g. PEPA: Actions are instantaneous and NOT associated with delays.

The remaining three rules encode the view that unhindered synchronisation (i.e. $\tau$ which is instantaneous) always takes precedence over delays. This is known as the *maximal progress* assumption. Note, that the hiding operator serves to delimit the scope of synchronisations, thereby allowing us to distinguish between potentially blocked, i.e. $a$, and definitely unhindered interactions, i.e. $\tau$.

Two consequences of these rules are particularly noteworthy:

First of all, IMC, in contrast to PEPA, does not define any way for delays to synchronise in a single exponential delay. Instead, IMC insists that delays interleave, thereby giving rise to compound delays the duration of which is phase-type distributed.

Second of all, non-determinism cannot always be completely resolved in a closed IMC. For this reason neither the action labelled transition system nor the Markov chain suffice as semantic models. This is the reason why IMC itself is also cast as a semantic model.

In general, proponents of IMC are very critical towards PEPA. As previously mentioned, the main point of criticism is that the class of exponential distributions is not closed under maximum. Indeed a central claim of IMC is that the phase type distribution arising from the interleaving of delays is *the* natural (and correct) characterisation of the maximum of the two involved exponential distributions. The price to pay for this increased correctness, however, is a significantly larger state space induced by the embedded phase-types as well as the aforementioned non-determinism, which forces us to use a more complicated semantic model.

Other stochastic process algebras include: TImed Processes for Performance evaluation (TIPP) [48], Markovian Process Algebra (MPA) [20], Stochastic $\pi$-calculus [87], Extended Markovian Process Algebra (EMPA) [21], BioAmbients [89], MoDeST [22], and StoKlaim (a stochastic extension of Klaim) [84].

Other stochastic modelling formalisms include: Stochastic Automata Networks (SAN) [85] and Stochastic Petri Nets (SPN) [81].

# Part IV

# Model Checking

## Chapter 6

# Logical Specification of Stochastic Properties

To verify a property of a system (regardless the type of system), we first need some way of expressing it. While we can always describe specific properties in a suitably rich mathematical logic — e.g. first order logic — we need to limit this expressiveness if we are to *automatically verify* any property that we can write down. More specifically, the challenge is to find a logic that is both expressive enough for the properties we are interested in, and admits efficient model checking algorithms.

For the qualitative analysis of concurrent systems, a number of temporal logics were developed with this aim. In practice the two most widely used are Computation Tree Logic (CTL) [42] and Linear Temporal Logic (LTL) [79], which are both subsets of the logic CTL*. The difference between CTL and LTL is in the treatment of time — an LTL formula refers to a specific execution path in the model, whereas a CTL formula refers to a tree of possible computations.

For a model with $N$ states and $M$ transitions, and a property $\Phi$, the complexity of CTL model checking is $O((N+M)|\Phi|)$, whereas LTL model checking is $O((N+M)2^{|\Phi|})$. On the other hand, it is generally considered to be more intuitive to specify properties in LTL [95]. As an example, consider the following formulae constructed from the 'next' ($\mathcal{X}$) and 'future' ($\mathcal{F}$) modalities (we will introduce these properly later in the chapter):

- (LTL) $\forall \mathcal{X}\mathcal{F}\,\Phi$: on all paths, after the current state there is a state in the future at which $\Phi$ holds.

- (LTL) $\forall \mathcal{F}\mathcal{X}\,\Phi$: on all paths, there is a state in the future for which the next state satisfies $\Phi$.

- (CTL) $\forall \mathcal{X} \, \forall \mathcal{F} \, \Phi$: on all paths out of all next states, there is a state in the future for which $\Phi$ holds.

- (CTL) $\forall \mathcal{F} \, \forall \mathcal{X} \, \Phi$: on all paths, there is a state in the future for which all successor states satisfy $\Phi$.

The first three of the above have identical semantics, however the last formula means something quite different — illustrating the care that must be taken when writing CTL formulae.

In the context of *qualitative* properties — relating to probability, time, and rewards — there have been a number of logics developed that extend CTL. The fact branching time was favoured over linear time is most likely because of the lower complexity of the model checking problem. The semantics and model checking algorithms of both CTL and LTL can naturally be extended to a probabilistic setting. Since there are an infinite number of computation paths in any non-trivial model, we need to assign probabilities to *measurable sets of paths*. In CTL, computation trees naturally form measurable sets of paths (so-called cylinder sets), and since either the satisfaction or violation of a given LTL formula can be demonstrated by a finite prefix of a path, we have a similar construction in both cases [13].

Both explicit state [63, 72] and symbolic [63] model checking algorithms have been extended to probabilistic systems. In the former, the graph reachability algorithms are essentially modified to solve *probabilistic* reachability problems. In the latter, Binary Decision Diagrams (BDDs) are extended to Multi-Terminal Binary Decision Diagrams (MTBDDs), which are efficient data structures for representing *real-valued* functions (i.e. those that map to a probability), rather than just Boolean functions.

In this chapter, we will describe the following logics, which are the ones most widely used in practice:

|                  | Discrete Time        | Continuous Time     |
|------------------|----------------------|---------------------|
| Without Rewards  | PCTL (Section 6.1)   | CSL (Section 6.3)   |
| With Rewards     | PRCTL (Section 6.2)  | CSRL (Section 6.4)  |

Although all of the above are branching time logics, the logic PCTL* [13] — a probabilistic extension of CTL* — has also been introduced, and is supported by the PRISM model checker [63]. As with CTL*, the complexity of PCTL* is exponential in the size of the formula. Note that it is doubly exponential, however, when there is non-determinism in the model. There has been some analogous work in the context of CSL, using regular expressions to describe path formulae [14, 15].

Before we describe each of the above logics in detail, let us introduce the basic syntax that is common to each. Importantly, there is a distinction between *state formulae* and *path formulae*, in that the former hold of states in the model, whereas the latter hold of (possibly infinite) sequences of states, or paths. We will use $\Phi$ to denote state formulae, and $\varphi$ to denote path formulae.

All the above logics contain the following state formulae in common, where $a \in AP$ is an atomic proposition, or label of a state:

$$\Phi ::= \mathtt{tt} \mid a \mid \Phi \wedge \Phi \mid \neg \Phi \mid \mathcal{P}_{\trianglelefteq p}(\varphi)$$

This consists of a propositional fragment, along with a probability measure formula. The latter states that the probability measure over paths that satisfy $\varphi$ is $\trianglelefteq p$, where $\trianglelefteq \in \{<, \leq, \geq, >\}$.

The common path formulae are as follows:

$$\varphi ::= \mathcal{X} \Phi \mid \Phi \mathcal{U} \Phi$$

$\mathcal{X} \Phi$ is the untimed next operator, which holds of a path if $\sigma$ if the next state $\sigma_1$ satisfies $\Phi$. $\Phi_1 \mathcal{U} \Phi_2$ is the untimed until operator, which holds of a path $\sigma$ if some state in the future satisfies $\Phi_2$, and all states before this point satisfy $\Phi_1$: $\exists i.\ \sigma_i \models \Phi_2 \wedge \forall j < i.\ \sigma_j \models \Phi_1$.

Note that various commonly-used operators can be derived as follows:

$$\begin{aligned}
\mathcal{F} \Phi &= \mathtt{tt} \mathcal{U} \Phi \\
\mathcal{G} \Phi &= \neg(\mathtt{tt} \mathcal{U} \neg \Phi) \\
\Phi_1 \mathcal{W} \Phi_2 &= (\mathcal{G} \Phi_1) \vee (\Phi_1 \mathcal{U} \Phi_2) \\
\Phi_1 \mathcal{R} \Phi_2 &= \neg(\neg \Phi_1 \mathcal{U} \neg \Phi_2)
\end{aligned}$$

$\mathcal{F} \Phi$ is the *future* operator, which states that $\Phi$ holds at some point in the future along a path. $\mathcal{G} \Phi$ is the *global* operator, which states that $\Phi$ holds for all states along a path. $\Phi_1 \mathcal{W} \Phi_2$ is the *weak until* operator, which states that $\Phi_1$ has to hold along the path until $\Phi_2$ holds (but $\Phi_2$ might not ever hold). Finally, $\Phi_1 \mathcal{R} \Phi_2$ is the *release* operator, which states that $\Phi_2$ must hold up to and including the point at which $\Phi_1$ becomes true, and if $\Phi_1$ never becomes true, then $\Phi_2$ must hold forever.

We will describe the semantics of this common subset of the logics (and subsequently, for each logic we introduce), in terms of DTMCs and CTMCs. A DTMC is a three-tuple $(S, \boldsymbol{P}, L)$, where $S$ is a non-empty finite set of states, $\boldsymbol{P} : S \times S \to [0, 1]$ is a stochastic matrix, and $L : S \times AP \to \{\mathtt{tt}, \mathtt{ff}\}$ is a labelling function. A CTMC is a four-tuple $(S, \boldsymbol{P}, r, L)$, where $r : S \to \mathbb{R}_{\geq 0}$ gives the exit rate of each state, and $S$, $\boldsymbol{P}$, and $L$ have the same meaning as for a DTMC.

A *path* $\sigma$ in a DTMC is a (possibly infinite) sequence of states $s_0, s_1, \ldots \in S$, such that for all $i < |\sigma| - 1$, $\boldsymbol{P}(s_i, s_{i+1}) > 0$. We write $\sigma[i] = s_i$, and $Paths(s)$ to be the set of paths such that $\sigma[0] = s$. A path $\sigma$ in a CTMC is a (possibly infinite) alternating sequence of states and durations $s_0, t_0, s_1, t_1, \ldots$, such that $s_i \in S$, $t_i \in \mathbb{R}_{>0}$, and for all $i < |\sigma| - 1$, $\boldsymbol{P}(s_i, s_{i+1}) > 0$ and $r(s_i) > 0$. As for a DTMC path, we write $\sigma[i] = s_i$, but we additionally define $\delta(\sigma, i) = t_i$ (the time spent in state $s_i$), and $\sigma@t = \sigma[i]$, where $i$ is the smallest index such that $t \leq \sum_{j=0}^{i} t_j$.

The semantics of the common subset of the logics is as follows. $s \models \Phi$ means that a state $s$ satisfies a state formula $\Phi$, and $\sigma \models \varphi$ means that a path $\sigma$ satisfies a path formula $\varphi$. Note that the semantics of the untimed next and until path formulae are the same for paths in both DTMCs and CTMCs, since there is no reference to time.

$$
\begin{aligned}
s &\models \mathtt{tt} & &\text{iff} & &\text{true} \\
s &\models a & &\text{iff} & &L(s, a) \\
s &\models \Phi_1 \wedge \Phi_2 & &\text{iff} & &s \models \Phi_1 \text{ and } s \models \Phi_2 \\
s &\models \neg \Phi & &\text{iff} & &s \not\models \Phi \\
s &\models \mathcal{P}_{\trianglelefteq p}(\varphi) & &\text{iff} & &\Pr\{\sigma \in Paths(s) \mid \sigma \models \varphi\} \trianglelefteq p \\
\sigma &\models \mathcal{X}\,\Phi & &\text{iff} & &\sigma[1] \models \Phi \\
\sigma &\models \Phi_1\,\mathcal{U}\,\Phi_2 & &\text{iff} & &\exists i.\ \sigma[i] \models \Phi_2 \text{ and } \forall j < i.\ \sigma[j] \models \Phi_1
\end{aligned}
$$

We will now describe each logic in turn. We will extend the above semantics in each case, but just describe the syntax, and the sort of properties that can be expressed. In Section 6.5, we will conclude with a brief discussion of uniformisation. In the next chapter, we will then relate these logics to the property specification languages of the PRISM and MRMC model checkers.

## 6.1 PCTL

Probabilistic Computation Tree Logic (PCTL) [53] is a logic for specifying properties of DTMCs and MDPs, and has the following syntax:

$$
\begin{aligned}
\Phi &::= \mathtt{tt} \mid a \mid \Phi \wedge \Phi \mid \neg \Phi \mid \mathcal{P}_{\trianglelefteq p}(\varphi) \mid \\
\varphi &::= \mathcal{X}\,\Phi \mid \Phi\,\mathcal{U}\,\Phi \mid \Phi\,\mathcal{U}^{\leq n}\,\Phi
\end{aligned}
$$

Here, we have introduced a *time-bounded* until operator. $\Phi_1\,\mathcal{U}^{\leq n}\,\Phi_2$ states that $\Phi_2$ will hold at some time $t \leq n$, and until that time, $\Phi_1$ will always hold. When $n = \infty$, this is the same as the unbounded until operator. More formally, the semantics is:

$$\sigma \models \Phi_1\,\mathcal{U}^{\leq n}\,\Phi_2 \text{ iff } \exists i \leq n.\ \sigma[i] \models \Phi_2 \text{ and } \forall j \leq i.\ \sigma[j] \models \Phi_1$$

In the MRMC model checker, the bounded until operator has been extended to take an arbitrary interval $I = [t_1, t_2]$, such that $t_1 \in \mathbb{N}_{\geq 0}$, $t_2 \in \mathbb{N}_{\geq 0} \cup \{\infty\}$, and $t_1 \leq t_2$. This has the semantics:

$$\sigma \models \Phi_1 \,\mathcal{U}^I\, \Phi_2 \text{ iff } \exists i \in I.\ \sigma[i] \models \Phi_2 \text{ and } \forall j \leq i.\ \sigma[j] \models \Phi_1$$

Explicit state model checking of PCTL can be performed by recursively checking the sub-terms in a formula. The only interesting case is the until operator, for which we need to calculate the probability of the formula holding in each state. For both the bounded and unbounded until operator, we first label the states that definitely satisfy the property ($\Phi_2$ holds), or definitely fail to satisfy the property (neither $\Phi_1$ nor $\Phi_2$ holds). The bounded until operator $\Phi_1 \,\mathcal{U}^{\leq n}\, \Phi_2$ then corresponds naïvely to raising to probability transition matrix of the DTMC to the power $n$, followed by a matrix-vector multiplication. We can improve the efficiency, however, using dynamic programming. The unbounded until operator corresponds to solving a set of linear equations.

### 6.1.1 Long-run properties in PCTL

In both PRISM and MRMC, PCTL has further been extended to allow long-run properties. More specifically, they support a long-run operator $\mathcal{L}_{\trianglelefteq p}(\Phi)$, which states that the probability of satisfying $\Phi$ in the steady state is $\trianglelefteq p$. Mathematically, a steady state distribution is only defined for an ergodic DTMC — specifically, one that is positive recurrent (for every state in the DTMC, the expected period before returning to it is finite) and aperiodic (it is not the case that any state can only return to itself in multiples of $k \geq 2$ time steps). The semantics of the long-run operator is slightly more general though (requiring only aperiodicity):

$$s \models \mathcal{L}_{\trianglelefteq p}(\Phi) \text{ iff } \lim_{n \to \infty} \Pr\{\,\sigma \in \mathit{Paths}(s) \mid \sigma[n] \models \Phi\,\} \trianglelefteq p$$

In practice, this means that the model checking algorithm consists of two stages. First, we identify the bottom strongly connected components (BSCCs) and compute the probability of reaching each BSCC from every transient state. We then compute the steady state distribution of each BSCC, which by definition must be ergodic (we require aperiodicity, and a BSCC is by definition irreducible, which implies positive recurrence when there are only finitely many states).

## 6.1.2 PCTL*

We can extend PCTL by allowing path formulae to be nested — in other words, allowing us to specify arbitrary LTL path formulae. This leads to the logic PCTL* [13]:

$$\begin{aligned} \Phi &::= \mathtt{tt} \mid a \mid \Phi \wedge \Phi \mid \neg \Phi \mid \mathcal{P}_{\trianglelefteq p}(\varphi) \\ \varphi &::= \Phi \mid \mathcal{X}\varphi \mid \varphi \mathcal{U} \varphi \mid \varphi \mathcal{U}^{\leq n} \varphi \mid \varphi \wedge \varphi \mid \neg \varphi \end{aligned}$$

Other than the nesting of path formulae, the only difference to PCTL in the above syntax is that we allow conjunction and negation of path formulae. The semantics of path formula is as follows, where we write $\sigma^i$ to be the suffix $\sigma[i], \sigma[i+1], \ldots$ of the path $\sigma$:

$$\begin{aligned} \sigma &\models \Phi & &\text{iff} & &\sigma[0] \models \Phi \\ \sigma &\models \mathcal{X}\varphi & &\text{iff} & &\sigma^1 \models \varphi \\ \sigma &\models \varphi_1 \mathcal{U} \varphi_2 & &\text{iff} & &\exists i.\ \sigma^i \models \varphi_2 \text{ and } \forall j < i.\ \sigma^j \models \varphi_1 \\ \sigma &\models \varphi_1 \mathcal{U}^{\leq n} \varphi_2 & &\text{iff} & &\exists i \leq n.\ \sigma^i \models \varphi_2 \text{ and } \forall j < i.\ \sigma^j \models \varphi_1 \\ \sigma &\models \varphi_1 \wedge \varphi_2 & &\text{iff} & &\sigma \models \varphi_1 \text{ and } \sigma \models \varphi_2 \\ \sigma &\models \neg \varphi & &\text{iff} & &\sigma \not\models \varphi \end{aligned}$$

PCTL* model checking is performed in a bottom-up recursive manner, in the same way as for PCTL. The difference is that when we reach a path probability operator, we convert the path formula into a Quantitative LTL Specification (QLS). This is a pair, $(\varphi, I)$, consisting of an LTL formula $\varphi$ (including the bounded until operator), and a probability interval of the form $[0, p]$ or $[p, 1]$, for $p \in [0, 1]$.

Verification of a QLS formula amounts to constructing an $\omega$-automaton corresponding to the path formula, and taking the product of this with the original system (a DTMC or MDP). This is often a Rabin automaton[1], since we require the $\omega$-automaton to be deterministic — deterministic Büchi automata do not accept all $\omega$-regular languages, whereas deterministic Rabin automata do. Alternatively, deterministic Streett[2] or Muller[3] automata are sometimes used. The QLS model checking problem is then reduced to probabilistic reachability (essentially, PCTL model checking) on the product automaton.

---

[1] The acceptance condition of a Rabin automaton is a set of pairs $(E_i, F_i)$ — a string is accepted if it results in a sequence of states where for some $i$, there is at least one state in $F_i$ that is visited infinitely often, and all states in $E_i$ are visited finitely often.

[2] The acceptance condition of a Streett automaton is the negation of the Rabin condition.

[3] The acceptance condition of a Muller automaton is a set of sets of states $F$ — a string is accepted if the set of states it visits infinitely often is an element of $F$. Rabin and Streett automata can be described as Muller automata.

Note that the complexity of PCTL* model checking is doubly exponential in the size of the formula — this can be reduced to singly exponential in the case of a DTMC, but not in the case of an MDP. PCTL* is supported by PRISM for both DTMCs and MDPs, but not by MRMC.

## 6.2 PRCTL

PCTL is used to specify properties of a DTMC, but we might want to build a reward structure on top of such a model. Rewards (or equivalently, costs) can be assigned to either states or transitions, or both. These can be used to capture a variety of metrics, including power consumption, monetary cost, and quality of service. To reason about such reward structures, Probabilistic Reward Computation Tree Logic (PRCTL) [11] was developed as an extension of PCTL. The syntax is as follows:

$$\begin{aligned}
\Phi &::= \ \texttt{tt} \ | \ a \ | \ \Phi \wedge \Phi \ | \ \neg\Phi \ | \ \mathcal{P}_{\triangleleft p}(\varphi) \ | \ \mathcal{L}_{\triangleleft p}(\Phi) \\
&\quad | \ \mathcal{E}_J(\Phi) \ | \ \mathcal{E}_J^n(\Phi) \ | \ \mathcal{C}_J^n(\Phi) \ | \ \mathcal{Y}_J^n(\Phi) \\
\varphi &::= \ \Phi \mathcal{U}_J^I \Phi
\end{aligned}$$

This is PCTL, extended with four additional state formulae, that allow us to reason about reward rates ($\mathcal{E}$), instantaneous rewards ($\mathcal{C}$), and accumulated rewards ($\mathcal{Y}$). The only change to the path formulae is the addition of a reward bound on the time-bounded until operator. $\Phi_1 \mathcal{U}_J^I \Phi_2$ means that $\Phi_2$ holds within a number of steps $n \in I$, that all states before this satisfy $\Phi_1$, and that the accumulated reward before satisfying $\Phi_2$ is in the interval $J$. Note that we use a superscript to talk about time, and a subscript to talk about rewards.

In the original paper [11], the only path formula considered is the time- and reward-bounded until operator, hence we only show this in the above. The other PCTL path formulae can also be included, however, and are supported by MRMC.

The reward operators are:

- $\mathcal{E}_J(\Phi)$ — the long-run expected reward per time unit, in states that satisfy $\Phi$, is within the interval $J$.

- $\mathcal{E}_J^n(\Phi)$ — the expected reward per time unit, in the first $n$ time steps, in states that satisfy $\Phi$, is within the interval $J$.

- $\mathcal{C}_J^n(\Phi)$ — the instantaneous reward at the $n$th time step, in states that satisfy $\Phi$, is within the interval $J$.

- $\mathcal{Y}_J^n(\Phi)$ — the expected accumulated reward until the $n$th transition, in states that satisfy $\Phi$, is within the interval $J$.

We interpret the semantics of PRCTL over a Discrete Markov Reward Model (DMRM). This is a 4-tuple $(S, \boldsymbol{P}, L, \rho)$, where $(S, \boldsymbol{P}, L)$ is a DTMC, and $\rho : S \to \mathbb{R}_{\geq 0}$ is a reward structure, which assigns a real value to each state that represents the reward accumulated when we leave the state (note that a self loop is interpreted as leaving and then re-entering the state). We can describe the PRCTL semantics more formally as follows:

$$
\begin{aligned}
s &\models \mathcal{L}_{\unlhd p}(\Phi) &&\text{iff} && \lim_{n\to\infty} \Pr\{\sigma \in \mathit{Paths}(s) \mid \sigma[n] \models \Phi\} \unlhd p \\
s &\models \mathcal{E}_J(\Phi) &&\text{iff} && g(s, \{s' \mid s' \models \Phi\}) \in J \\
s &\models \mathcal{E}_J^n(\Phi) &&\text{iff} && g(s, \{s' \mid s' \models \Phi\}, n) \in J \\
s &\models \mathcal{C}_J^n(\Phi) &&\text{iff} && \rho(s, \{s' \mid s' \models \Phi\}, n) \in J \\
s &\models \mathcal{Y}_J^n(\Phi) &&\text{iff} && y(s, \{s' \mid s' \models \Phi\}, n) \in J \\
\sigma &\models \Phi_1 \mathcal{U}_J^I \Phi_2 &&\text{iff} && \exists i \in I.\ \sigma[i] \models \Phi_2 \text{ and } \forall i < j.\ \sigma[i] \models \Phi_1 \\
& && && \text{and } \sum_{j=0}^{i-1} \rho(\sigma[j]) \in J
\end{aligned}
$$

where we make use of the following reward measures:

$$
\begin{aligned}
g(s, S', n) &= \frac{1}{n+1} \sum_{i=0, \sigma_s[i] \in S'}^{n} \mathbb{E}(\rho(\sigma_s[i])) \\
g(s, S') &= \lim_{n \to \infty} g(s, S', n) \\
\rho(s, S', n) &= \sum_{s' \in S'} \rho(s') \pi(s, s', n) \\
y(s, S', n) &= \sum_{i=0}^{n-1} \rho(s, S', i) \\
\pi(s, s', n) &= \sum_{t \in S} \pi(s, t, i) \cdot \pi(t, s', n-i) \quad \text{for } 0 \leq i \leq n
\end{aligned}
$$

Here, $\pi(s, s', n)$ denotes the probability of being in state $s'$ after $n$ steps, given that we start in state $s$, and its definition comes directly from the Chapman-Kolmogorov equations.

The PRCTL reward operators are supported directly in MRMC, whereas PRISM has a slightly different syntax for reward-based properties. We will discuss this in the next chapter.

## 6.3 CSL

Both PCTL and PRCTL are discrete-time logics, used in the context of DTMCs and MDPs. We can move to continuous time with only minor changes in the logics. The biggest difference comes in the model checking algorithms themselves, in that uniformisation-based techniques are employed. Continuous Stochastic Logic (CSL) [12,16] is a logic for expressing properties of CTMCs, and has the following syntax:

$$\begin{aligned} \Phi &::= \texttt{tt} \mid a \mid \Phi \wedge \Phi \mid \neg \Phi \mid \mathcal{P}_{\trianglelefteq p}(\varphi) \mid \mathcal{S}_{\trianglelefteq p}(\Phi) \\ \varphi &::= \mathcal{X}\,\Phi \mid \Phi\,\mathcal{U}\,\Phi \mid \Phi\,\mathcal{U}^I\,\Phi \end{aligned}$$

The main difference between PCTL and CSL is that since the latter is a continuous time logic, the intervals on the path operators can be real-valued. The operator $\mathcal{S}$ is a steady state operator, and is just the continuous time analogue of the long-run operator $\mathcal{L}$ of PRCTL.

As an example of the sort of properties we can express, consider the following CSL formula, for $AP = \{\,Error,\,Completed\,\}$:

$$\mathcal{P}_{\geq 0.9}(\neg Error\,\mathcal{U}^{[0,10]}\,Completed)$$

This will be satisfied by all states from which there is a probability of at least 0.9 that we will reach a '*Completed*' state within 10 time units, without encountering any '*Error*' states before that point. Note that the unit of time is implicit to the model, and is only relevant with respect to our interpretation of the results.

More formally, the semantics of the new CSL operators are as follows:

$$\begin{aligned} s &\models \mathcal{S}_{\trianglelefteq p}(\Phi) &&\text{iff}&& \lim_{t\to\infty}\Pr\{\,\sigma \in Paths(s) \mid \sigma@t \models \Phi\,\} \trianglelefteq p \\ \sigma &\models \Phi_1\,\mathcal{U}^I\,\Phi_2 &&\text{iff}&& \exists t \in I.\ \sigma@t \models \Phi_2 \text{ and } \forall t' < t.\ \sigma@t' \models \Phi_1 \end{aligned}$$

In addition to the above operators, as presented in [16], we can additionally define a *time-bounded next operator* [18]. The path formula $\mathcal{X}^I\,\Phi$ requires that the next state satisfies $\Phi$, *and* that the transition will take place in the time interval $I$. Such an operator does not make sense in the context of PCTL, where transitions occur at discrete points in time (hence their duration is abstract). The semantics of the timed next operator is as follows:

$$\sigma \models \mathcal{X}^I\,\Phi \text{ iff } \sigma[1] \models \Phi \text{ and } \delta(\sigma,0) \in I$$

The standard model checking algorithm for the CSL time-bounded until operator is based upon uniformisation — we have to consider the three cases of $[0,t]$, $[t,\infty]$, and $[t_1,t_2]$ for the time interval, but the algorithm is essentially a first passage time analysis [17]. For the timed next operator, the model checking algorithm boils down to a matrix-vector multiplication. Note that the untimed next and until operators can be model checked as per PCTL by considering the embedded DTMC.

One problem we face with CTMCs is that uniformisation does *not* preserve the validity of all CSL formulae. This is a problem if we perform lumpability based abstractions, since these are typically based on a uniformised CTMC [73]. We will write CSL\X to mean the subset of CSL without the next operator. If two CTMCs are weakly bisimilar, then the validity of all CSL\X formulae is preserved [18]. A consequence is that the uniformisation of a CTMC preserves CSL\X equivalence.

### 6.3.1 CSL Variants

Path-based reward variables are regular expressions (expressed as finite state automata) that allow us to reason about more complex behaviours than with the standard CSL path operators [68]. This idea has subsequently been explored in a logical fashion, by allowing regular expressions in place of path formulae.

The first such variant of CSL that was proposed was pathCSL [14], in which path formulae are time-bounded regular expressions, concerning sequences of pairs of state formulae and action types. This assumes that the transitions on the CTMC are labelled with an action type $a \in Act$, for some finite set $Act$:

$$\begin{array}{rcl} \Phi & ::= & \mathtt{tt} \mid a \mid \Phi \wedge \Phi \mid \neg \Phi \mid \mathcal{P}_{\trianglelefteq p}(\varphi) \mid \mathcal{S}_{\trianglelefteq p}(\Phi) \\ \varphi & ::= & \langle \alpha \rangle^{\leq t} \Phi \\ \alpha & ::= & \epsilon \mid \Phi a \mid \alpha\alpha \mid \alpha + \alpha \mid \alpha^* \end{array}$$

Here, $\alpha$ is a regular expression over the alphabet of pairs of state formulae and action types — $\Sigma = \{ \Phi a \mid \Phi \text{ is a state formula}, a \in Act \}$.

Path formulae are interpreted over finite paths $\sigma = s_0, a_0, \ldots, s_n, a_{n-1}, s_n$, which consist of alternating states $s_i \in S$ and actions $a_i \in Act$. As previously, $\sigma@t$ is the state of $\sigma$ at time $t$, and $\delta(\sigma, i)$ is the residence time in state $s_i$. We write $\sigma(t_1, t_2)$ to mean the fragment of the path between times $t_1$ and $t_2$ inclusive, and $c(\sigma)$ to be the completion time for the path $\sigma$ (i.e. the total time taken for the path). The semantics of the path formula $\langle \alpha \rangle^{\leq t} \Phi$ is as follows:

$$\sigma \models \langle \alpha \rangle^{\leq t} \Phi \text{ iff } \exists t' \in [0, t].\ \sigma@t' \models \Phi \text{ and } \sigma(0, t') \in Paths(\alpha)$$

Here, the set $Paths(\alpha)$ of all paths that satisfy the regular expression $\alpha$ is defined recursively as follows:

$$
\begin{array}{rcll}
\sigma & \in & Paths(\epsilon) & \text{iff} \quad \sigma = s \in S \\
\sigma & \in & Paths(\Phi a) & \text{iff} \quad \sigma = s_0, a, s_1 \text{ and } s_0 \models \Phi \\
\sigma & \in & Paths(\alpha_1 \alpha_2) & \text{iff} \quad \exists t' \in [0, c(\sigma)].\ \sigma(0, t') \in Paths(\alpha_1) \\
& & & \qquad \text{and } \sigma(t', c(\sigma)) \in Paths(\alpha_2) \\
\sigma & \in & Paths(\alpha_1 + \alpha_2) & \text{iff} \quad \sigma \in Paths(\alpha_1) \text{ or } \sigma \in Paths(\alpha_2) \\
\sigma & \in & Paths(\alpha^*) & \text{iff} \quad \exists i \in \mathbb{N}.\ \sigma \in Paths(\alpha^i)
\end{array}
$$

pathCSL was later extended to asCSL (CSL with actions and state labels) [15], which is essentially the same logic, but with minor syntactic changes. Again, transitions on the CTMC are labelled with an action type, but properties can refer to actions types $a \in Act \cup \{\surd\}$, where $\surd \notin Act$. The pseudo-action $\surd$ is always immediately executable and does not change the state of the CTMC — its main use is at the end of a regular expression, allowing us to write $(\Phi, \surd)$ to require the final state to satisfy $\Phi$, without requiring any particular action type afterwards (or indeed, any subsequent transition at all).

The syntax of asCSL is as follows:

$$
\begin{array}{rcl}
\Phi & ::= & \texttt{tt} \mid a \mid \Phi \wedge \Phi \mid \neg\Phi \mid \mathcal{P}_{\trianglelefteq p}(\varphi) \mid \mathcal{S}_{\trianglelefteq p}(\Phi) \\
\varphi & ::= & \alpha^I \\
\alpha & ::= & \epsilon \mid (\Phi, a) \mid \alpha; \alpha \mid \alpha \cup \alpha \mid \alpha^*
\end{array}
$$

The semantics of path formulae is as follows:

$\sigma \models \alpha^I$ iff there exists a finite prefix $\sigma'$ of $\sigma$ such that
$$\sigma \in Paths(\alpha) \text{ and } c(\sigma) \in I$$

$Paths(\alpha)$ is defined recursively as before (we denote by $\sigma_i$ the prefix of $\sigma$ up to state $s_i$, and by $\sigma^i$ the suffix of $\sigma$ from state $s_i$ onwards):

$$
\begin{array}{rcll}
\sigma & \in & Paths(\epsilon) & \text{iff} \quad |\sigma| = 0 \\
\sigma & \in & Paths(\Phi, a) & \text{iff} \quad \sigma = s_0, a, s_1 \text{ and } s_0 \models \Phi \text{ and } \delta(\sigma, 0) > 0 \\
\sigma & \in & Paths(\Phi, \surd) & \text{iff} \quad \sigma = s \text{ and } s \models \Phi \\
\sigma & \in & Paths(\alpha_1; \alpha_2) & \text{iff} \quad \exists i \leq |\sigma|.\ \sigma_i \in Paths(\alpha_1) \text{ and } \sigma^i \in Paths(\alpha_2) \\
\sigma & \in & Paths(\alpha_1 + \alpha_2) & \text{iff} \quad \sigma \in Paths(\alpha_1) \cup Paths(\alpha_2) \\
\sigma & \in & Paths(\alpha^*) & \text{iff} \quad \exists i \in \mathbb{N}.\ \sigma \in Paths(\alpha^i)
\end{array}
$$

These ideas have not yet been implemented in the PRISM or MRMC model checkers, however the use of regular expressions for transient analysis has been developed and implemented in the context of PEPA [31]. Here, they are called *stochastic probes* — a regular expression is written, that specifies a sequence of actions to observe in the PEPA model. The question we then ask is: what is the first passage time until we reach a state where we have observed a sequence of actions that is in the language described by this regular expression? A common example of such a property is in measuring the response time of a component in the model.

A stochastic probe is compiled into a PEPA component that corresponds to the regular expression, in much the same way as the automata-theoretic approach to LTL model checking. Importantly, the component only contains passive activities, and so it does not alter the behaviour of the original model (it only adds additional information by introducing additional states). The problem then reduces to a first passage time analysis. This is supported by the PEPA plug-in for Eclipse [94].

Note that both pathCSL and asCSL are in a sense less expressive than a hypothetical CSL*, in which the CSL path formulae can be nested. This is because, in the above, a single time bound is given for the entire path. Hence model checking can be performed via an automata product construction, similar to LTL model checking, followed by a time-bounded reachability analysis, or first passage time analysis[4]. As an example of the difference, consider a path formula $\mathcal{X}^{[0,1]} \mathcal{X}^{[0,1]} \Phi$, which states that we perform two transitions, each within one time unit, to reach a state satisfying $\Phi$. This is clearly different to $(\mathcal{X} \mathcal{X} \Phi)^{[0,2]}$, expressible in asCSL, which states that we perform two transitions within two time units, to reach a state satisfying $\Phi$. The first formula is stricter, since a smaller set of paths will satisfy it.

## 6.4  CSRL

In the same way as with discrete time models, we can add reward structures to CTMCs. The main difference is that the reward assigned to a state is not a fixed reward that is given for each time step we occupy that state, but a *rate* of reward acquisition.

To reason about reward-structured CTMCs, the Continuous Stochastic Reward Logic (CSRL) [33] was developed as an extension of CSL. It has the following syntax:

$$\begin{aligned} \Phi &::= \texttt{tt} \mid a \mid \Phi \wedge \Phi \mid \neg \Phi \mid \mathcal{P}_{\trianglelefteq p}(\varphi) \mid \mathcal{S}_{\trianglelefteq p}(\Phi) \\ \varphi &::= \mathcal{X}_J^I \Phi \mid \Phi \mathcal{U}_J^I \Phi \end{aligned}$$

---

[4]Note that the computation of a first passage time and a time to absorption is the same, if we make the target state absorbing.

The only extension to CSL is the addition of the time- and reward-bounded next and until operators. $\mathcal{X}_J^I \Phi$ states that the next state satisfies $\Phi$, and the transition is made at some time $t \in I$, and the accumulated reward until time $t$ is in the interval $J$. The until operator $\Phi_1 \mathcal{U}_J^I \Phi_2$ is the continuous-time analogue of the corresponding PRCTL formula.

The semantics of CSRL is interpreted over a Continuous time Markov Reward Model (CMRM). This is a tuple $(S, \boldsymbol{P}, r, L, \rho, \iota)$, where $(S, \boldsymbol{P}, r, L)$ is a CTMC, $\rho : S \to \mathbb{R}_{\geq 0}$ is a reward structure describing the *rate* per time unit at which a reward is accumulated in each state, and $\iota : S \times S \to \mathbb{R}_{\geq 0}$ is a reward structure describing the *impulse reward* accumulated when a transition is made between two states. It must be the case that for all $s \in S$, $\iota(s, s) = 0$. We define the accumulated reward along a path $\sigma$ at time $t$ to be:

$$y_\sigma(t) = \rho(\sigma@t) . \left( t - \sum_{j=0}^{i-1} \delta(\sigma, j) \right) + \sum_{j=0}^{i-1} \rho(\sigma[j]) . \delta(\sigma, j) + \sum_{j=0}^{i-1} \iota(\sigma[j], \sigma[j+1])$$

The semantics of the new operators in CSRL is then as follows:

$$
\begin{aligned}
s &\models \mathcal{S}_{\trianglelefteq p}(\Phi) &&\text{iff} \quad \lim_{t \to \infty} \Pr\{\sigma \in \mathit{Paths}(s) \mid \sigma@t \models \Phi\} \trianglelefteq p \\
\sigma &\models \mathcal{X}_J^I \Phi &&\text{iff} \quad \sigma[1] \models \Phi \text{ and } \delta(\sigma, 0) \in I \text{ and } y_\sigma(\delta(\sigma, 0)) \in J \\
\sigma &\models \Phi_1 \mathcal{U}_J^I \Phi_2 &&\text{iff} \quad \exists t \in I.\ \sigma@t \models \Phi_2 \text{ and } \forall t' < t.\ \sigma@t' \models \Phi_1 \\
& && \quad \text{and } y_\sigma(t) \in J
\end{aligned}
$$

There is no analogue in CSRL of the PRCTL $\mathcal{E}$, $\mathcal{C}$, and $\mathcal{Y}$ state formulae, which means that we cannot reason about the long-run expected rate of a reward, the expected reward up to time $t$, the instantaneous rate of reward at time $t$, or the expected accumulated reward until time $t$. We can still express accumulated rewards over *individual paths*, however, thanks to the reward-bounded path operators. MRMC supports CSRL directly, whereas PRISM uses its own reward operators, which are the same for both continuous and discrete time properties — they allow us to talk about reachability, cumulative, instantaneous, and steady state rewards [6]. We will describe these in detail in the next chapter.

## 6.5  A Note on Uniformisation

The heart of the model checking algorithms for computing time-bounded probabilistic reachability properties in CTMCs are based on uniformisation. Recall that we previously presented a CTMC as a four-tuple $(S, \boldsymbol{P}, r, L)$, where $S$ is the state space, $\boldsymbol{P}$ describes the transition probability between each pair of states, $r$ describes the exit rate of each state, and $L$ is a labelling function. If we assign a numerical index to each state $s \in S$, then we can write $\boldsymbol{P}$ as a *stochastic matrix* — i.e. the rows all sum to one.

An alternative way of characterising a CTMC is in terms of an *infinitesimal generator matrix* $\boldsymbol{Q}$. If we consider a CTMC with $N$ states, such that $\boldsymbol{P}$ is its probabilistic transition matrix (an $N \times N$ stochastic matrix), and $r(i) = r(s)$ describes the exit rate of state $s$ with index $0 \leq i < N$, then we can define the elements of $\boldsymbol{Q}$ as follows:

$$\begin{aligned} \boldsymbol{Q}(i,j) &= r(i)\boldsymbol{P}(i,j) & \text{if } i \neq j \\ \boldsymbol{Q}(i,i) &= -r(i)\sum_{j \neq i} \boldsymbol{P}(i,j) \end{aligned} \quad (6.1)$$

If $r(i)$ is a constant rate $\lambda$ for all $i$, then the CTMC is said to be *uniformised*, and we can write $\boldsymbol{Q}$ in the following form:

$$\boldsymbol{Q} = \lambda(\boldsymbol{P} - \boldsymbol{I})$$

In general, this is not the case, but given $\boldsymbol{Q}$ as defined in Equation 6.1, and a unformisation constant $\lambda$, we can construct a uniformised probability transition matrix $\overline{\boldsymbol{P}}$ as follows:

$$\overline{\boldsymbol{P}} = \frac{\boldsymbol{Q}}{\lambda} + \boldsymbol{I}$$

We require that $\lambda \geq \lambda_{min} = \max_i |\boldsymbol{Q}(i,i)|$ in order for $\overline{\boldsymbol{P}}$ to be a stochastic matrix.

For transient analysis, as used by the model checking algorithms to compute probabilistic reachability properties (a first passage time analysis), it is best to choose the smallest value of $\lambda$ possible. However, we need to be careful if we want to ensure that the embedded DTMC $\overline{\boldsymbol{P}}$ of the uniformised CTMC preserved the ergodicity of $\boldsymbol{Q}$. If we are not careful, it is possible for $\overline{\boldsymbol{P}}$ to be *cyclic*, such that it does not converge on a steady state distribution for all initial distributions over the states. The PRISM model checker sets the value of $\lambda$ to be slightly greater than $\lambda_{min}$.

It is interesting to note, however, that under different circumstances — depending on what the purpose of the unformisation is — a different choice of $\lambda$ may be optimal. One example of this is in the construction of stochastic bounds, where we try to find a DTMC whose steady state solution is an upper bound on that of the original DTMC, using a stochastic ordering [93]. The usual idea is to find an upper bound that exhibits some property such as *lumpability* [75], so that its state space can be reduced in size. Since stochastic bounds are formulated in terms of DTMCs, we need to apply them to the embedded DTMC of a uniformised CTMC — hence we need to choose an appropriate unformisation constant $\lambda$.

It has been shown in [37] that the optimal choice of $\lambda$ in this circumstance is $2\lambda_{min}$. Intuitively, this introduces additional probability mass to the diagonal of $\overline{P}$, which gives more flexibility when constructing a stochastic bound, allowing the bound to be tighter.

# Chapter 7

# Stochastic Model Checking

In this chapter we will look at the capabilities of two of the leading probabilistic model checkers. We look at PRISM in Section 7.1 and MRMC in Section 7.2.

## 7.1 PRISM (version 3.3.1)

The probabilistic symbolic model checker PRISM [7,78] was developed at the University of Birmingham by Marta Kwiatkowska's group and is currently being maintained and developed by the same research group at Oxford University. The core team includes Marta Kwiatkowska, Gethin Norman and Dave Parker, and they have been developing PRISM since 1999. PRISM provides support for DTMCs, CTMCs, and MDPs, using a probabilistic guarded-command language (known as the PRISM language) as its primary interface[1].

An overview of the structure of the PRISM tool, taken from [77], is shown in Figure 7.1. The model checking algorithms are based around three engines:

- The *sparse engine* is based around a sparse matrix representations of the model, and corresponds to explicit state model checking.

- The *MTBDD engine* is based on using multi-terminal binary decision diagrams[2] to represent the model, and corresponds to probabilistic symbolic model checking (this is where the name PRISM came from).

---

[1]PRISM also supports PEPA as an input language, but only active-passive synchronisation is implemented. A PEPA model maps onto a CTMC.

[2]An MTBDD is like a BDD, except that the terminal nodes represent probabilities (in this context), rather than truth values. If there are only a few probability values that an expression can evaluate to, this can lead to an efficient representation.

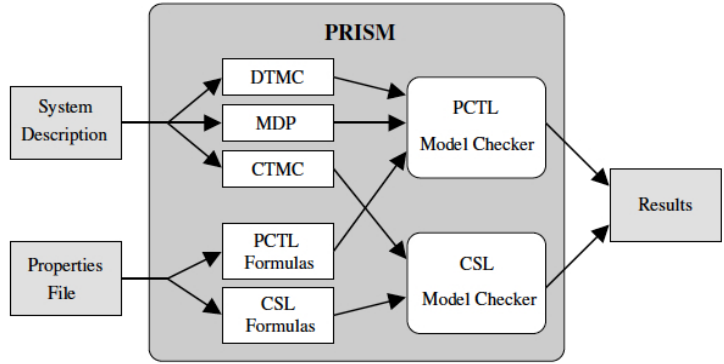

Figure 7.1: Overview of the structure of PRISM

MTBDDs can be used to efficiently represent very large models that have a regular structure, but this is not the case for all models, and it depends on the specific ordering of variables in the PRISM model — in other words, the order in which variables are declared in the PRISM file can affect the size of the MTBDD representation of its transition matrix.

- The *hybrid engine* uses a combination of the above two representations [76]. The transition matrix of the model is stored in an MTBDD, whereas the iteration vector — recording the probability of each state satisfying a given property — is stored as a full array. This gives better results than the MTBDD engine for most models, since the majority of states have different probabilities of a satisfying a given property, meaning that the iteration vector cannot be stored efficiently as an MTBDD. Because of this, the hybrid engine is selected in PRISM by default.

We will give an overview of the PRISM language in Section 7.1.1, which maps onto a DTMC, CTMC, or MDP, and may be enriched with reward structures. We will then describe the PRISM property specification language in Section 7.1.2. The following logics are supported, for each type of model:

|      | PCTL | PCTL* | CSL | Rewards |
|------|------|-------|-----|---------|
| DTMC | ✓    | ✓     |     | ✓       |
| MDP  | ✓    | ✓     |     | ✓[†]    |
| CTMC |      |       | ✓   | ✓       |

[†]Only certain reward properties (reachability and instantaneous rewards) are supported by MDPs.

Note that PRISM does not support PRCTL and CSRL as described in the previous chapter. Instead, it has its own syntax for reward specification, which we will describe in Section 7.1.2. PRISM does support CTL model checking in the context of MDPs, where the CTL property $A\varphi$ has the same semantics as the PCTL property $\mathcal{P}_{\geq 1}(\varphi)$. This is because $\varphi$ must hold over all adversaries (even unfair ones) in order for the probability to lie in the interval $[1, 1]$, and therefore be $\geq 1$. These two properties also have the the same semantics for a DTMC, since Zeno paths have probability zero. Hence the only sensible interpretation of $A\varphi$ on a DTMC is that $\varphi$ almost certainly holds.

### 7.1.1 The PRISM Language

The PRISM modelling language [6] is a simple, state-based language based on the Reactive Modules formalism [10]. It has two main components: *modules* and *variables*. A PRISM model consists of a number of modules that run in parallel. Each contains local variables that constitute its state — which may only take on a fixed range of values — and a set of *guarded commands* that describe its behaviour. Global variables are also allowed, which can be read and modified by all modules[3].

As an example, consider a PRISM module with only two local variables, `x` and `y`, declared as follows:

```
x :   [0..3] init 0;
y :   bool init false;
```

`x` is an integer variable, taking a value in the set $\{0, 1, 2, 3\}$, and `y` is a Boolean variable. There are eight possible states of the module (not all may be reachable), and we specify a state by giving the values of all the variables in the module. The initial state is given by the initial values of the variables, as specified in the declaration — in this case, $(\texttt{x} = 0, \texttt{y} = \text{false})$. We can specify a *set* of states in a compact way, by giving an inequality on the values of certain variables. For example, the condition $\texttt{x} \geq 2$ describes all the states where `x` has a value of 2 or 3, and `y` has any value (there are four such states).

The behaviour of a PRISM module is described by a set of guarded commands. A guarded command has the following general form, and is prefixed with an optional action name $A$:

$$[A]G \to p_1 : U_1 + \cdots + p_n : U_n$$

---

[3]Commands that synchronise are not allowed to modify global variables, to prevent race conditions.

The guard $G$ is a condition on the state of the variables (i.e. it determines a set of states in which the command can execute) — both local to the module and in other modules. If $G$ is true, then with probability $p_i$, update $U_i$ is performed. It must hold that $\sum_i p_i = 1$, and the probabilities must be specified — if the sum contains only one element, the probability can be omitted, and will be taken to be one. An update specifies how the state of the local variables changes (we write $x$ to refer to the old state of a variable and $x'$ to refer to the new state). If we are specifying a continuous time model, rates will be used in place of probabilities.

In a DTMC or CTMC model, there must only be at most one guard that evaluates to true for each state of the module. In an MDP model, multiple guards can evaluate to true, which introduces local non-determinism.

We build a PRISM model, known as a *system*, by specifying how to compose its modules. The following operators are supported, based on those in the Communicating Sequential Processes (CSP) process algebra [64]:

- $M_1 || M_2$ — parallel composition of modules $M_1$ and $M_2$, synchronising on all action names that appear in both $M_1$ and $M_2$.

- $M_1 ||| M_2$ — asynchronous parallel composition of modules $M_1$ and $M_2$.

- $M_1 |[L]| M_2$ — parallel composition of modules $M_1$ and $M_2$, synchronising only on action names in $L$.

- $M/L$ — hiding[4] of action names in $L$, in module $M$. The module behaves as $M$, except that action names in $L$ are renamed to the internal action name $\tau$.

- $M/\rho$ — renaming of action name $a$ in module $M$ to $\rho(a)$.

When we compose two modules, any commands prefixed by an action name that they synchronise over must execute together. For discrete time models, we take the joint probability distribution over the updates to perform after executing the two commands. In continuous time models, we similarly multiply each pair of rates together.

PRISM uses an interleaved model of concurrency. This means that when two modules run in parallel, and can perform commands independently, there is a question as to the order of their interleaving. In an MDP model, this is treated as a non-deterministic choice, whereas for a DTMC model, the choice must be probabilistic. In PRISM, if their are $n$ commands concurrently enabled, then the probability of executing each command is $\frac{1}{n}$. Intuitively, we can think of a probabilistic choice being made as to which command to execute, followed by the probabilistic choice specified in the command. This is illustrated in Figure 7.2, where the PRISM modules are shown graphically, and the enabled command on each has only one possible choice.

---

[4]Note that this is hiding in the style of CSP, as opposed to *restriction* in the style of CCS. The latter blocks communication on names appearing in $L$.

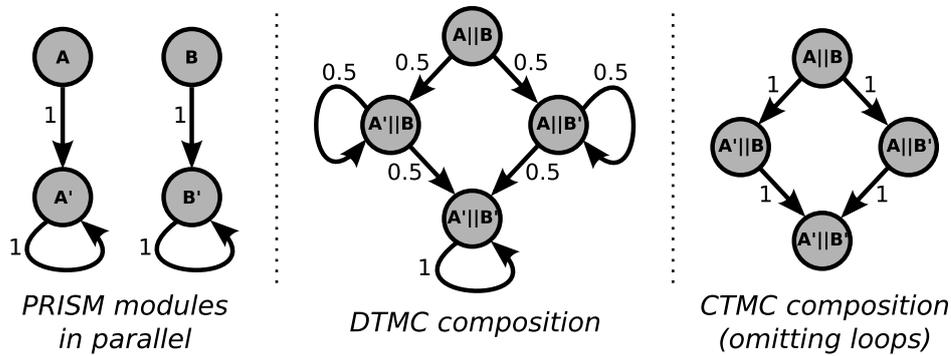

Figure 7.2: Composition of PRISM modules

For a CTMC model, the commands take place at a particular *rate*, rather than with a certain probability. When PRISM takes the composition of the modules, every enabled command in a state of the system is allowed to proceed at the specified rate. Intuitively, this means that the exit rate from a state of the system is the *sum* of the exit rates of the individual states, and the probability of one command proceeding over another depends on the relative exit rate. This is illustrated again in Figure 7.2, where the labels on the transitions should be interpreted as rates rather than probabilities. We have omitted the self loops, as they do not change the behaviour of the CTMC.

The final feature of PRISM that we will talk about here is its reward structures. Each model can be given multiple reward structures, and each reward structure is defined separately to the rest of the model (i.e. outside any module definitions). A reward structure is simply a list of pairs of states and real values, corresponding to the reward (or cost) for that state. In practice, we specify such a structure compactly, by describing a *set* of states, and giving an expression for the reward in each case, in terms of the state variables. A simple example of a reward structure that only depends on one variable (`x : [0..10]`) is:

```
rewards
    x < 5 :  20 * x;
    x >= 5 :  100;
endrewards
```

This assigns a reward of 20x to all states in which x has a value less than 5, and a reward of 100 to all the other states (when the value of x is between 5 and 10 inclusive).

## 7.1.2 Property Specification

Properties in PRISM are based on the two main logics we described in the previous section — PCTL for discrete time models, and CSL for continuous time models. The atomic properties are just sets of states, which can be defined by constraints on the variables in the model, in the same way as when specifying reward structures. We can also define *labels*, which can subsequently be used as a shorthand for such sets of states.

There are three operators that PRISM provides for state formulae — P for the probability of satisfying a path formula, S for the long-run or steady state probability of being in a certain set of states, and R for reward properties. The P and S operators are much the same as in PCTL and CSL, and have the following syntax, where $\Phi$ is a state formula and $\varphi$ is a path formula:

$$\begin{aligned} &\texttt{P } \trianglelefteq r \texttt{ [ } \varphi \texttt{ ]} \\ &\texttt{P =? [ } \varphi \texttt{ ]} \\ &\texttt{S } \trianglelefteq r \texttt{ [ } \Phi \texttt{ ]} \\ &\texttt{S =? [ } \Phi \texttt{ ]} \end{aligned}$$

The only difference is the addition of the *qualitative properties*, P =? and S =?, which evaluate to a probability. This has no impact on the model checking algorithms, since we have to compute the probabilities in any case, but is useful from a practical standpoint. In the case of an MDP, we need to replace '=?' with one of 'min=?' or 'max=?', depending on whether we want the minimum or maximum probability of the property holding. PRISM supports all the path formulae of PCTL, PCTL*, and CSL, including the derived CTL operators for convenience.

Reward properties in PRISM are a little different to those in PRCTL and CSRL that we saw in the previous chapter:

$$\begin{aligned} &\texttt{R } \trianglelefteq r \texttt{ [ } R_{prop} \texttt{ ]} \\ &\texttt{R =? [ } R_{prop} \texttt{ ]} \\ &\texttt{R \{ } \textit{reward structure} \texttt{ \} } \trianglelefteq r \texttt{ [ } R_{prop} \texttt{ ]} \\ &\texttt{R \{ } \textit{reward structure} \texttt{ \} =? [ } R_{prop} \texttt{ ]} \end{aligned}$$

The information in braces is required for models with multiple reward structures — this can either be the name of the reward structure, or an index (with 1 being the first reward structure, etc.). For the purposes of experimentation, an undefined integer constant can used in place of a fixed index.

As before, in the case of an MDP, we replace '=?' by 'min=?' or 'max=?', depending on whether we are interested in the minimum or maximum value of the reward.

There are four types of reward property supported by PRISM:

- `F` $\Phi$ — a 'reachability reward'. This specifies the expected reward accumulated along a path until a state satisfying $\Phi$ is reached. If we allow quantitative properties in PRCTL (i.e. of the form $\mathcal{P}_{=?}(\varphi)$), then we can express the property `R <= `$r$` [ F `$\Phi$` ]` as follows:

$$\frac{\mathcal{P}_{=?}(\texttt{tt}\,\mathcal{U}_{[0,r]}\,\Phi)}{\mathcal{P}_{=?}(\texttt{tt}\,\mathcal{U}\,\Phi)}$$

- `C <= `$t$ — a 'cumulative reward'. This specifies the expected reward accumulated along a path until a time $t$ (which is either a natural number or a real number, depending on whether we are verifying a discrete- or continuous-time model). `R <= `$r$` [ C <= `$t$` ]` is equivalent to the PRCTL property $\mathcal{Y}^t_{[0,r]}(\texttt{tt})$.

- `I = `$t$ — an 'instantaneous reward'. This specifies the expected reward (or rate of reward in the case of a continuous-time model) of the model at a particular time $t$. `R <= `$r$` [ I = `$t$` ]` is equivalent to the PRCTL property $\mathcal{C}^t_{[0,r]}(\texttt{tt})$.

- `S` — a 'steady state reward'. This specifies the reward per time unit in the long run. `R <= `$r$` [ S ]` is equivalent to the PRCTL property $\mathcal{E}_{[0,r]}(\texttt{tt})$.

All of the above are supported for DTMCs and CTMCs, but only reachability rewards and instantaneous rewards are supported for MDPs, and only by the sparse and MTBDD engines (not the hybrid engine).

The property 'R <= $r$ [ F $\Phi$ ]' is quite different from the PRCTL path formula $\texttt{tt}\,\mathcal{U}^{[0,\infty]}_{[0,r]}\,\Phi$. The former is a state formula, requiring the *expectation* (over all paths) of the accumulated reward until $\Phi$ holds to be $\leq r$. The latter, on the contrary, is a path formula, requiring the accumulated reward on a specific path to be $\leq r$. This means that in PRCTL we can impose a reward bound on *every* path that satisfies a certain path formula, but we cannot impose a bound on the expected reward over all paths that satisfy the formula. The opposite situation is true of PRISM reward properties.

Note that in PRCTL, the reward operators are qualified with a state formula $\Phi$ — for example, $\mathcal{C}^N_J(\Phi)$ for instantaneous rewards. This means that we only include any contributions to the reward from those states that satisfy $\Phi$, when we verify the property. There is no equivalent to this in PRISM, but if we know in advance which states satisfy $\Phi$, then we could manually encode this as a separate reward structure that only assigns a non-zero reward to these states.

All of the above properties we have discussed are state formulae — that is to say, the model checking problem only makes sense if we say *which* state we are interested in. When we ask PRISM to verify a property, it makes the following assumptions:

- If the property is *Boolean*, e.g. 'P < 0.01 [ F "error"]', then PRISM will return true iff the property holds of *all* states in the model. If we are only interested in a particular set of states, we can use a logical implication. For example:

$$\text{"state\_x"} \Rightarrow \text{P < 0.01 [ F "error"]}$$

- If the property is *quantitative*, e.g. 'P =? [ F "error"]', then PRISM will return the value of the property for the *initial* state of the model. If we are interested in a particular state, we can use a *filter*, by identifying the state inside the quantitative operator. For example:

$$\text{P =? [ F "error" \{"state\_x"\}]}$$

If we specify a set of states in this way, we also need to say whether we are interested in the minimum or maximum value of the property in this set. For example:

$$\text{P =? [ F "error" \{"state\_x\_or\_y"\} \{max\} ]}$$

Note that there is a difference between the above '{max}' notation for the maximum value of a property over a set of states, and the 'max=?' notation for the maximum value of a property over all schedulers. For example, in the context of an MDP, we could write the following property:

$$\text{P max=? [ F "error" \{"state\_x\_or\_y"\} \{min\} ]}$$

This allows us to talk about the smallest probability of there being an error in the future, starting from a certain set of states ("state_x_or_y"), given a worst-case adversary.

### 7.1.3 Simulation in PRISM

In addition to stochastic model checking, PRISM supports *discrete event simulation* of models. There are two interfaces through which this can be performed:

1. *Simulation of a single execution of the model.* This can be thought of a debugging engine, in which we can either automatically simulate a certain number of steps (using a random number generator for probabilistic choices), or manually select the next state. The interface also supports backtracking, and allows us to explore the state space of the model.

2. *Approximate probabilistic model checking* [54]. Given a quantitative property (of the form `P =?` or `R =?`), the model is simulated multiple times, and an approximate value for the property is given by the expected value of the property over all runs. We can specify the desired precision — i.e. that the actual value lies within $\epsilon$ of the approximate value with a given confidence interval — and PRISM will compute the required number of samples to ensure this.

### 7.1.4 Experiments in PRISM

One additional feature provided by PRISM is its support for so-called *experiments*. The idea is that we often want to verify not just one property, but a class of properties where one parameter is varied. For example, we might want to know the probability of a path formula holding as we change the value of some variable, or as we change the time bound of a path operator.

To perform experiments, we simply declare some constants in the PRISM model, but leave them undefined. When we come to specify a property, we can then use these variables, and PRISM will prompt us for the values to use. This is typically specified as a start and end value, and a step size, indicating a range of values. PRISM will then model check the property for each value of the variables.

Note that the experiments feature can make use of either the stochastic model checker or the simulator. In both cases, we perform a parameter sweep such that we either verify the same property for a set of different models, or a set of different properties for the same model. The intent is often to plot a graph of the probability or reward value of a quantitative property with respect to the parameter we vary. Note that experimentation and simulation are two different and complimentary features.

## 7.2 MRMC (version 1.4.1)

The Markov Reward Model Checker (MRMC) [72] is an explicit state stochastic model checker, whose focus has historically been on reward-structured Markovian models. It supports five different modelling formalisms: DTMCs, CTMCs, discrete time Markov reward models (DMRMs), continuous time Markov reward models (CMRMs), and continuous time Markov decision processes (CTMDPs). Technically, MRMC supports CTMDPIs, which are CTMDPs that contain *internal* non-determinism in addition to external non-determinism that is controlled by an environment.

Unlike PRISM, the MRMC tool does not accept a modelling language directly — instead, it takes an explicit description of the state space, transition system, state labels, and reward structures of the model as input. This is given by number of simple text files that explicitly list every state in the model, and every transition between states, etc. As a command line tool, it is designed to be used a back-end to other systems, in which higher level compositional modelling languages are used — for example, PEPA and the PRISM language.

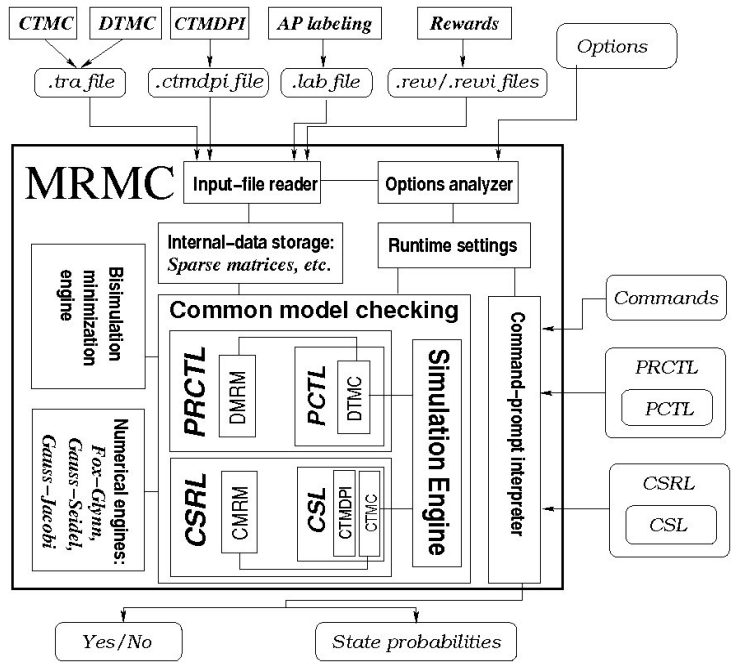

Figure 7.3: Overview of the structure of MRMC

A diagram of the structure of MRMC, taken from [5], is shown in Figure 7.3. There are two fundamental model checking engines:

- *Explicit-state stochastic model checking* — this uses the model checking algorithms that we mentioned briefly in the previous chapter. Unlike PRISM, there is no use of symbolic model checking, but bisimulation minimisation techniques are used to reduce the size of the state space.

- *Statistical model checking* — this uses discrete event simulation to validate a property, returning confidence bounds on its correctness. There is also a hybrid mode for steady state properties, in which the probabilities of reaching bottom strongly-connected components are computed numerically, but the steady state behaviour is explored by simulation.

MRMC supports the following logics for each modelling formalism:

|        | PCTL | PRCTL | CSL | CSRL |
|--------|------|-------|-----|------|
| DTMC   | ✓    |       |     |      |
| DMRM   |      | ✓     |     |      |
| CTMC   |      |       | ✓   |      |
| CMRM   |      |       |     | ✓    |
| CTMDPI |      |       | ✓†  |      |

†For CTMDPIs, only time-bounded reachability properties are supported.

We will not describe the syntax for property specification here, as it is essentially the same as that of the logics described in the previous chapter.

## 7.3 Other Model Checking Tools

In this section, we present a brief description of other stochastic model checking tools.

### 7.3.1 RAPTURE (version 2.0.0)

The RAPTURE tool [8, 69] provides support for verifying quantified reachability properties over a Probabilistic Transition System (PTS). A PTS is similar to an MDP, in that given a state and an action label, there is a probabilistic choice as to the next state to transition to. Unlike an MDP, however, it is *reactive*, and two PTSs can perform a CSP-style synchronisation over shared actions. In this sense, it is more similar to the process algebra IMC, described in Section 5.2.

In RAPTURE, a system is specified in terms of a number of channels and processes, as well as a set of a global initial states and a set of global final states. The latter describes the set of states that we are interested in reaching, and so can be thought of as specifying the reachability property as part of the system. In addition to calculating such reachability probabilities, RAPTURE employs a number of reduction techniques in order to mitigate the problem of state space explosion.

### 7.3.2 PASS

The probabilistic model checker *PASS (Predicate Abstraction for Stochastic Systems)* can be used to analyse concurrent probabilistic programs which map to DTMCs or MDPs with infinite states. It is based on predicate abstraction and automatic abstraction refinement. Models are specified in a variant of the PRISM language, which allows infinite data types.

Since PASS does not unfold the entire state space of the original model, the tool is not restricted to finite models — unlike, for example, Rapture [35] and the magnifying-lens abstraction of [9]. Instead, PASS takes the approach of counterexample-guided abstraction refinement (CEGAR), which uses predicate abstraction to maintain a *finite* abstract model. Analysis of this abstract model is typically very efficient since it has a small state space. Importantly, the analysis is *safe*, in that it yields probability intervals that are guaranteed to contain the probabilities of the corresponding properties in the original model. The size of the interval is used to quantify the approximation error caused by the abstraction. If this error is too large, the abstraction is refined — using diagnostic information (effectively, a counterexample) obtained from the abstract model.

This process is described in [56, 96]. Note that a major difference to conventional CEGAR for predicate abstraction is that the counterexamples are *Markov chains*, rather than single paths. The tool makes use of SMT solvers for computing finite abstractions, numerical methods for computing probabilities on these abstractions, and interpolation as part of the abstraction refinement mechanism. PASS has been successfully applied to network protocols, and serves as a test platform for different refinement methods.

### 7.3.3 PARAM

*PARAM* is a tool for handling parametric variants of models specified in a variant of the PRISM language. It extends the PRISM language with the possibility of defining unknown parameters, and later using such parameters to specify probability distributions. PARAM is capable of computing unbounded reachability probabilities for parametric DTMCs.

To solve this problem, Daws [36] devised a language-theoretic approach, in which the transition probabilities are taken to be letters in an alphabet. In this way, the model can be viewed as a finite state automaton. Given this, a regular expression describing the language of the automaton is computed, using the *state elimination* [66] method. The regular expression is then recursively evaluated, which results in a rational function over the parameters of the model. Gruber and Johannsen [50] have shown, however, that the size of the regular expression of a finite automaton explodes — for an automaton with $n$ states, it has size $n^{\Theta(\log n)}$.

The core of PARAM is also rooted in the state elimination algorithm. The key difference to [36] is that instead of post-processing a (possibly prohibitively large) regular expression, the state elimination and rational function computation stages are intertwined. More precisely, regular expressions are not used in the state elimination step — instead, the edges are labelled directly with the appropriate rational function to represent the flow of probabilities. This also means that the process remains in the domain of Markov chains, rather than working on a finite automaton representation.

In addition to DTMCs, PARAM can also handle special classes of MDPs, as well as reachability rewards for Markov reward models. To speed up computations, it computes the (strong and weak) bisimulation quotient of the parametric model.

### 7.3.4 INFAMY

*INFAMY* is a tool for model checking CSL formulae on infinite state CTMCs, which are specified in a variant of the PRISM language. It implements the first CSL model checking algorithm to use *truncation*. This algorithm enables the automatic analysis of infinite (or very large) CTMCs, and, unlike [90,91], it can be applied to arbitrarily structured (finite or infinite) CTMC models. Given a CSL property, the algorithm proceeds in two phases:

1. A finite truncation depth is computed, which is sufficient to check the property up to a given accuracy.

2. The property is verified on the resulting truncated model, using the CSL model checking algorithm for finite CTMCs [17,74].

Note that computing a sufficient truncation depth for CSL model checking is more challenging than for transient analysis, since the required depth depends not only on the desired precision and the characteristics of the model — formulae involving until operators and nested sub-formulae must also be handled. Nevertheless, results for transient analysis form an important building block of the algorithm.

INFAMY provides a number of methods for finding a stopping criterion for the state-space exploration. Currently, the supported methods are Uniform, Layered, FSP, and FSP exp — for a comparison, see [51, 52]. It is interesting to observe the tradeoff between the time at which the state space exploration is stopped, and the memory needed to store the finite truncation of the state space. In addition to CSL properties, INFAMY can also verify certain reward properties.

# Part V

# Performance Evaluation

## Chapter 8

# Tools for Performance Evaluation

In this chapter, we will give an overview of the tool support for the performance evaluation techniques that we described in the previous chapter. Since there is a certain degree of overlap between performance evaluation and stochastic model checking, there are many tools that offer similar functionality to one another. Our focus in this chapter, however, will be on the tool support for analysing performance properties that are *not* expressed in a logic.

Of the many performance evaluation tools that are available, we will focus our attention on three in particular:

1. *PEPA Tools* (Section 8.1). We introduced the Performance Evaluation Process Algebra (PEPA) in Section 5.1, and we will give, in this chapter, a survey of the main tools that are available for analysing PEPA models.

2. *Möbius* (Section 8.2). This is a popular tool that supports multiple high-level modelling formalisms, which can be composed to form hierarchical models. There are a number of analysis engines, based around numerical solution of Markov chains, and discrete event simulation.

3. *MATLAB* (Section 8.3). Many mathematicians write specific models directly in the MATLAB programming language, which provides in-built support for matrices and standard mathematical operations.

## 8.1 PEPA Tools

There has been a long history of tool support for PEPA, initially based around numerically computing the steady state distribution of the underlying CTMC, but subsequently encompassing first passage time analysis, stochastic simulation, fluid-flow approximation and model checking of Continuous Stochastic Logic (CSL). The original PEPA tool was the PEPA Workbench [46], which was developed in 1994 as two independent versions — one written in SML, and one in Java. Over time, the Java tool gained dominance, with a number of researchers and masters students extending its functionality. However, due to the involvement of many people, and the lack of a structured design, it eventually became bloated and too difficult to maintain.

In 2006, the PEPA plug-in project [94] began, as an attempt to reimplement the PEPA workbench using good design principles. Rather than a stand-alone application, it was decided to develop a plug-in for the Eclipse platform [1], both to make use of a standard interface, and to allow easier integration with other tools. In addition to numerical solution of models, the PEPA plug-in supports fluid-flow approximation [60] and stochastic simulation [23] of PEPA models. More recently, support for compositional abstraction and CSL model checking has been added [92].

In parallel to the PEPA workbench and the PEPA plug-in, tools have been developed for transient analysis of PEPA models. In particular, the Imperial PEPA Compiler is a tool that compiles a PEPA model into the input language of HYDRA [24], allowing distributed computation of response time distributions. Stochastic probes [31] are used to specify the response time to measure. Recently, the Imperial PEPA Compiler has been superseded by the International PEPA Compiler (IPC) [2], which can be used both as a stand-alone application and through Eclipse — allowing integration with the PEPA plug-in.

In addition to these language-specific tools, PEPA is also supported by a number of more general purpose performance evaluation and stochastic model checking tools. A subset of PEPA (allowing only active-passive synchronisation) is directly supported by the PRISM model checker [63] (see Section 7.1), and Möbius [34] supports an extension of PEPA, called $PEPA_k$, which we will describe in Section 8.2.

A summary of the four main tools for PEPA, and their functionality, is shown in Table 8.1. Note that we have not labelled PRISM as supporting stochastic simulation, because it does not support simulation-based performance evaluation in the same way as the PEPA plug-in and Möbius do. Currently, PRISM only supports discrete event simulation in the context of debugging and statistical model checking.

|  | PEPA Plug-in | IPC | PRISM | Möbius |
|---|:---:|:---:|:---:|:---:|
| Steady State Solution | ✓ | ✓ | ✓ | ✓ |
| Stochastic Probes |  | ✓ |  |  |
| CSL Model Checking | ✓ |  | ✓ |  |
| Stochastic Simulation | ✓ |  |  | ✓ |
| Fluid-Flow Approximation | ✓ |  |  |  |

Table 8.1: Performance evaluation features of the main PEPA tools

We will discuss stochastic probes in more detail in Section 8.1.1, and population-level analysis using stochastic simulation and fluid-flow approximation in Section 8.1.2.

## 8.1.1 Transient Analysis

The International PEPA compiler (IPC) [2] is a tool for computing passage time distributions for PEPA models. It supports an extension of PEPA that allows immediate actions (i.e. actions that are instantaneously taken), functional rates [61], and arrays of processes (which all cooperate over the same set of action types). Note that transitions corresponding to immediate actions are always given priority over stochastic transitions[1].

To illustrate the idea of a passage time query, let us consider a simple PEPA sequential component. This corresponds to a client, sending a request to a server and waiting for a response:

$$\begin{aligned} Client &\stackrel{def}{=} (request, r).Client' \\ Client' &\stackrel{def}{=} (response, \top).Client \end{aligned}$$

This could potentially be composed with a complex model of a server, which processes the request in multiple stages, or has to contend for some shared resource such as a remote database. From the point of view of the client, however, we might like to ask a simple question — "after I make a request, how long do I have to wait before I receive a response?"

To ask such queries, IPC supports a language called eXtended Stochastic Probe (XSP) [31]. The basic idea is to specify when we *start* measuring the passage time, and when we *stop* measuring it. In the case of our example query, we start when we observe a *request* activity, and we *stop* when we

---

[1]This corresponds to the maximal progress assumption in Interval Markov Chains (IMC). See Section 5.2.

observe a *response* activity. The stochastic probe would therefore be as follows:

$$Client :: request : \text{start}, response : \text{stop}$$

The meaning of *Client* :: *P* is to *attach* the probe *P* to the *Client* component. This is important if we want to measure the response time of an individual client — there may be other components in the model that also perform *request* and *response* activities with the server, but we are only interested in those activities that involve the client.

In general, we might want to ask much more complex queries than the above, and so it is not sufficient to identify a single pair of activities that cause us to start and stop measuring the passage time. Instead, we are able to specify *regular expressions* of activities[2]. An example of such a stochastic probe is as follows:

$$Client :: (request_1 \mid request_2) : \text{start}, (response_A, response_B) : \text{stop}$$

This specifies the passage time between observing either a $request_1$ or a $request_2$ activity, and observing a $response_A$ activity followed by a $response_B$ activity.

Given a PEPA model and a stochastic probe, IPC compiles the probe into a PEPA component that records the state of the probe (i.e. whether it is measuring or not measuring the passage time), but does not affect the state of the model. This is composed with the original PEPA model, and may cause its state space to increase — depending on the complexity of the regular expressions in the probe. The approach is similar to the automata-theoretic approach to LTL model checking, which we discussed briefly in Section 6.1.2.

At the backend of IPC is the ipclib/HYDRA toolchain [24]. HYDRA, the HYpergraph-based Distributed Response time Analyser [38], is a tool for computing first passage time distributions of a CTMC, using uniformisation[3]. It does so in a distributed fashion, using a hypergraph partitioning of the state space — each processor is assigned a set of states in the CTMC, in such a way that the required communication between processors is minimised.

---

[2]XSP also allows us to use *state specification* as guards — that is to say, we only observe an activity if the current state of the model satisfies a certain property. We will not describe this here, but refer the reader to [31].

[3]HYDRA is an extension of DNAmaca, which is a tool for solving the steady state distribution of large Markov chains.

### 8.1.2 Population-Level Analysis

In PEPA, we often want to write a model that consists of a certain number of identical components in parallel. For example, if we want to model a system where four clients connect to two servers, we could write a system equation that looks something like the following:

$$(Client \parallel Client \parallel Client \parallel Client) \bowtie_{\{request, response\}} (Server \parallel Server)$$

To make it easier to specify such models, an *aggregation combinator* was introduced into the language. If we want to represent $n$ copies of a sequential process $P$ in parallel (with no shared activities between them), we can simply write $P[n]$. This means that the above system equation can be simplified to:

$$Client[4] \bowtie_{\{request, response\}} Server[2]$$

The problem with models of this form is that the underlying state space can become intractably large even for relatively small numbers of components. Let us consider a term $P[n]$, where the sequential process $P$ has $m$ different configurations ($|\text{ds}(P)| = m$). If we naïvely generate the state space of the underlying CTMC, we will find that it contains $m^n$ states. In other words, the state space grows exponentially large with respect to number of components.

We can be cleverer than this if we notice that two components in the same state are indistinguishable from one another — more precisely, they are strongly bisimilar. For example, we cannot distinguish between $P_1 \parallel P_2 \parallel P_1$ and $P_1 \parallel P_1 \parallel P_2$, since observably they both correspond to two processes in state $P_1$ and one process in state $P_2$. We can therefore *aggregate* such states in the CTMC induced by the PEPA model, since they exhibit (ordinary) *lumpability*. In general, the number of states in the aggregated process $P[n]$ will be[4]:

$$\binom{n+m-1}{n} = \frac{(n+m-1)!}{n!(m-1)!}$$

Whilst this is a significant improvement, the number of states still grows quickly with $n$ — for a fixed $m$, the number of states will grow as a polynomial in $n$ of degree $m-1$.

To combat this, the PEPA plug-in supports two techniques[5]:

---

[4]Combinatorically, this is the number of ways we can place $n$ indistinguishable balls into $m$ distinguishable urns.

[5]Initially, these analyses were performed by compiling PEPA into the input language for Dizzy [88], a chemical kinetics simulation package. They are now, however, supported natively by the PEPA plug-in.

1. *Stochastic simulation* [23]. We simulate the CTMC induced by the PEPA model using Gillespie's algorithm [45] (and more recent improvements on it). The essential idea is a discrete-event simulation of a *reaction-based* representation of the CTMC.

2. *Fluid-flow approximation* [60]. Rather than having a discrete state space that describes the number of components in each state, we approximate this by a *continuous state space*. This leads to an alternative semantics for PEPA, which describes the model as a system of ordinary differential equations (ODEs). In particular, there is one differential equation for each state of a sequential component, but the number of equations does not grow with the number of components.

Both of these approaches take a *population-level* view of the PEPA model. In other words, we use PEPA to model the behaviour of the system at the level of an individual component, or species, and then use this to analyse the emergent behaviour of the population. In doing so, we lose information about an individual — for example, we can no longer ask how long it takes for an individual client to receive a response from a server, but we *can* ask questions about how the *number* of clients waiting for a response changes over time.

## 8.2   Möbius (version 2.3)

Möbius [34] is a multi-paradigm performance evaluation tool. It supports multiple modelling languages, which can be used to model individual components of a system, and then be combined using common notions of composition. A model in Möbius essentially consists of a number of *states*, along with various *actions* that allow it to change state. Actions have a duration, but unlike in most stochastic process algebras, they can be *generally distributed*[6].

There are two main types of analysis in Möbius, and the analysis engines are common to all the modelling formalisms:

1. *Distributed discrete-event simulation* — this can be used to analyse both the transient and steady state behaviour of a model. It supports distribution of simulation runs to multiple machines over the network, and automatically collects and collates the results (this is done by automatic remote login to other machines via `rsh` or `ssh`).

---

[6]Möbius has in-built support for the binomial, deterministic, gamma, exponential, Erlang, beta, hyper-exponential, negative binomial, geometric, uniform, triangular, Weibull, conditional Weibull, normal, and log-normal distributions.

2. *Numerical solution* — if a model contains only actions with exponentially distributed durations, then it can be solved numerically — both for transient and steady state analysis. If a model contains both exponential and deterministic transitions, and only one deterministic transition is enabled at any time (whose duration does not depend on the state of the model), then its steady state can be numerically solved. Numerical solution, however, requires the model to be small enough to fit in memory.

Models in Möbius are inherently *hierarchical* — individual components of the system are modelled separately, and then composed to build a new model. This can itself be a sub-model of a larger composed model. Composition is primarily based around having *shared state* between sub-models, although *actions* can alternatively be shared. There are three ways of specifying composition in Möbius:

1. *Replicate/join* — the composition is described by a tree, whose leaf nodes are individual components. A parent node is either a *replicate-k* node, which has a single child and corresponds to instantiating $k$ copies of that child, or a *join* node, which has multiple children and composes them. Both replicate and join nodes must specify which variables are shared between their children, and which are local.

2. *Graph composition* — this is an alternative to replicate/join, where there is no requirement that the graph forms a tree, and there is no replicate node. This allows us to specify in a more intuitive way the composition of more than two sub-models that share certain elements of one another's state. Note however that any graph composition can be expressed as a replicate/join, and vice versa.

3. *Action synchronisation* — rather than sharing state between sub-models, we can share their *actions*. The enabling conditions of a shared action are the union of its enabling conditions in each of the sub-models, and the new rate of the action can be specified by the user. Like replicate/join, an action synchronisation is described as a tree.

There are four main modelling formalisms supported by Möbius — stochastic activity networks, buckets and balls, PEPA, and fault trees. These are described in detail in the Möbius user manual [4], but we will give an overview here for completeness.

### 8.2.1   Stochastic Activity Networks (SANs)

Stochastic Activity Networks (SANs)[7] [82] are similar to stochastic Petri nets (SPNs) [81], in that they are a graphical formalism. A SAN consists of four primitive elements:

- A *place* corresponds to a state of the model, and can contain a number of *tokens*. The number of tokens assigned to a place is called its *marking*.

- An *activity* is an action, linking a set of input places to a set of output places. Activities can be either timed or instantaneous.

- An *input gate* is a Boolean function on the markings of the input places of an activity, and must evaluate to true in order for the activity to be enabled.

- An *output gate* is a function that defines the marking changes that occur in an output place when an activity fires.

### 8.2.2   Buckets and Balls

Like SANs, buckets and balls are a graphical formalism, but are much simpler. These are used to model components where the capabilities of an SAN are not needed (e.g. the use of input and output gates). There are two primitive elements:

- A *bucket* corresponds to a state in the model, and can contain a number of *balls*.

- An *transition* corresponds to an event that transfers balls between two buckets, and is drawn as a directed edge. The cardinality of a transition determines how many balls will be transferred, and it can only fire if there are sufficient balls in the source bucket. As with SAN activities, transitions can be either timed or instantaneous.

### 8.2.3   PEPA$_k$

PEPA$_k$ is an extension of the PEPA language (described in Section 5.1), which adds the following language features:

---

[7]Note that these should not be confused with Stochastic Automata Networks [85], which have the same acronym, but are very different.

- Sequential process definitions can be given *parameters*, which correspond to natural numbers.

- Activities can be prefixed by *guards*, which are Boolean predicates on the parameters. An activity

An example of a $\text{PEPA}_k$ component is the following, where $P[a]$ is defined for all $a \geq 0$:
$$\begin{aligned} P[a] &= [a > 0] \Rightarrow (\alpha, r).P[a-1] \\ &+ [a = 0] \Rightarrow (\alpha, r).P[a] \end{aligned}$$

Note that we can always translate a $\text{PEPA}_k$ model into PEPA, by expanding out the sequential process definitions, starting from the system equation. There is no syntactic guarantee that the model will be finite, however.

### 8.2.4 Fault Trees

Fault trees [39] are a formalism used to analyse the reliability of a system, by analysing the relationship between the failure of individual components, and of the entire system. The basic idea is that a system failure is described as a logical combination of component failures — Möbius also supports a subset of *dynamic* fault trees, allowing failures to also depend on the *sequence* in which components fail.

The elements of a fault tree can be either active, or inactive. There are three types of element:

- A *node* is a failure state of the system. The root element of a fault tree must always be a node, but nodes can also be present at intermediate levels of the tree.

- An *event* is a failure of a component in the system. Events form the leaf elements of the fault tree, and correspond to activities in the other Möbius formalisms.

- A *logic gate* connects a number of elements of the fault tree (its inputs) to a parent element (its output), and is triggered by a condition. Logic gates may be static (AND, OR, XOR, and $K$-of-N[8]), or dynamic (priority AND). The output state of a static gate depends only on the current state of its inputs, whereas that of a dynamic gate may also depend on the *past* state of its inputs. Möbius only supports one dynamic gate — the priority AND, whose output becomes active if and only if its inputs become active in a certain order, specified by the user.

---
[8]The output of a $K$-of-N gate is active if and only if $K$ of its inputs are active.

## 8.3　MATLAB

Rather than using tools that support language-based or graphical formalisms, mathematicians that build stochastic models tend to work at a lower level, using *parameterisation* to compactly specify a model. To this end, MATLAB [3] is a widely used programming language, since it natively supports vectors and matrices, and various standard mathematical operations. Whilst this lacks the compositionality of higher level language-based formalisms, it is in many ways closer to the parametric specification of models preferred by mathematicians.